\documentclass[useAMS,usenatbib,fleqn]{mn2e}

\usepackage{amssymb}
\usepackage{times}
\usepackage{graphicx}
\usepackage{epstopdf}
\usepackage{subfigure}
\usepackage[T1]{fontenc}
\usepackage{aecompl} 
\usepackage{multirow}
\usepackage{pdflscape}
\usepackage{rotating}

\usepackage[fleqn]{amsmath}
\usepackage{mathrsfs}
\usepackage{xcolor}

\usepackage{anyfontsize}

\usepackage[breaklinks=true]{hyperref}
\hypersetup{colorlinks=true,citecolor = blue}
\usepackage{cleveref}

\usepackage{scalerel}
\usepackage{tikz}
\usetikzlibrary{svg.path}

\newcommand{\orcidpng}[1]{\href{https://orcid.org/#15}{\includegraphics[scale=0.1]{/Users/maurosereno/Documents/bozze/ORCID-iD_icon-128x128.png}}}

\definecolor{orcidlogocol}{HTML}{A6CE39}
\tikzset{
  orcidlogo/.pic={
    \fill[orcidlogocol] svg{M256,128c0,70.7-57.3,128-128,128C57.3,256,0,198.7,0,128C0,57.3,57.3,0,128,0C198.7,0,256,57.3,256,128z};
    \fill[white] svg{M86.3,186.2H70.9V79.1h15.4v48.4V186.2z}
                 svg{M108.9,79.1h41.6c39.6,0,57,28.3,57,53.6c0,27.5-21.5,53.6-56.8,53.6h-41.8V79.1z M124.3,172.4h24.5c34.9,0,42.9-26.5,42.9-39.7c0-21.5-13.7-39.7-43.7-39.7h-23.7V172.4z}
                 svg{M88.7,56.8c0,5.5-4.5,10.1-10.1,10.1c-5.6,0-10.1-4.6-10.1-10.1c0-5.6,4.5-10.1,10.1-10.1C84.2,46.7,88.7,51.3,88.7,56.8z};
  }
}

\newcommand\orcid[1]{\href{https://orcid.org/#1}{\mbox{\scalerel*{
\begin{tikzpicture}[yscale=-1,transform shape]
\pic{orcidlogo};
\end{tikzpicture}
}{|}}}}

\newcommand{\beq}{\begin{equation}}
\newcommand{\eeq}{\end{equation}}

\def\gs{\mathrel{\lower0.6ex\hbox{$\buildrel {\textstyle >}\over{\scriptstyle \sim}$}}}
\def\ls{\mathrel{\lower0.6ex\hbox{$\buildrel {\textstyle <}\over{\scriptstyle \sim}$}}}
\newcommand{\simgt}{\lower.5ex\hbox{$\; \buildrel > \over \sim \;$}}
\newcommand{\simlt}{\lower.5ex\hbox{$\; \buildrel < \over \sim \;$}}

\newcommand{\aap}{A\&A}

\newcommand{\apj}{ApJ}

\newcommand{\apjs}{ApJS}
\newcommand{\aj}{AJ}

\newcommand{\jcap}{J. Cosmol. Astropart. Phys.}
\newcommand{\pasj}{PASJ}

\newcommand{\mnras}{MNRAS}

\newcommand{\ssr}{Space Science Reviews}

\defcitealias{se+et15_comalit_I}{CoMaLit-I}
\defcitealias{ser+al15_comalit_II}{CoMaLit-II}
\defcitealias{ser15_comalit_III}{CoMaLit-III}
\defcitealias{se+et15_comalit_IV}{CoMaLit-IV}
\defcitealias{se+et17_comalit_V}{CoMaLit-V}
\defcitealias{xxl_I_pie+al16}{XXL~Paper~I}
\defcitealias{xxl_II_pac+al16}{XXL~Paper~II}
\defcitealias{xxl_III_gil+al16}{XXL~Paper~III}
\defcitealias{xxl_IV_lie+al16}{XXL~Paper~IV}
\defcitealias{xxl_XIII_eck+al16}{XXL~Paper~XIII}
\defcitealias{xxl_XX_ada+al18}{XXL~Paper~XX}
\defcitealias{xxl_XXV_pac+al18}{XXL~Paper~XXV}
\defcitealias{xxl_XXXVIII_ser+al19}{XXL~Paper~XXXVIII}

\begin{document}

\title[CoMaLit -- VI]{CoMaLit -- VI. Intrinsic scatter in stacked relations. The weak lensing AMICO galaxy clusters in KiDS-DR3}
\author[Sereno et al.]{Mauro Sereno \orcid{0000-0003-0302-0325},$^{1,2}$\thanks{E-mail: mauro.sereno@inaf.it (MS)}
Stefano Ettori \orcid{0000-0003-4117-8617},$^{1,2}$ 
Giorgio F. Lesci,$^{3}$
Federico Marulli \orcid{0000-0002-8850-0303},$^{3,1,2}$
\newauthor 
Matteo Maturi,$^{4,5}$
Lauro Moscardini\orcid{0000-0002-3473-6716},$^{3,1,2}$
Mario Radovich,$^{6}$
Fabio Bellagamba,$^{1,3}$
\newauthor 
Mauro Roncarelli$^{1,3}$
\\
$^1$INAF -- Osservatorio di Astrofisica e Scienza dello Spazio di Bologna, via Piero Gobetti 93/3, I-40129 Bologna, Italy\\
$^2$INFN, Sezione di Bologna, viale Berti Pichat 6/2, 40127 Bologna, Italy\\
$^3$Dipartimento di Fisica e Astronomia, Alma  Mater Studiorum Universit\`a di Bologna, via Piero Gobetti 93/2, I-40129 Bologna, Italy\\
$^4$Zentrum f\"ur Astronomie, Universit\"at Heidelberg, Philosophenweg 12, D-69120 Heidelberg, Germany\\
$^5$ITP, Universit\"at Heidelberg, Philosophenweg 16, 69120 Heidelberg, Germany\\
$^6$INAF -- Osservatorio Astronomico di Padova, vicolo dell’Osservatorio 5, 35122 Padova, Italy
}


\maketitle

\begin{abstract}
Unbiased and precise mass calibration of galaxy clusters is crucial to fully exploit galaxy clusters as cosmological probes. Stacking of weak lensing signal allows us to measure observable--mass relations down to less massive halos halos without extrapolation. 
We propose a Bayesian inference method to constrain the intrinsic scatter of the mass proxy in stacked analyses. The scatter of the stacked data is rescaled with respect to the individual scatter based on the number of binned clusters. We apply this method to the galaxy clusters detected with the AMICO (Adaptive Matched Identifier of Clustered Objects) algorithm in the third data release of the Kilo-Degree Survey. The results confirm the optical richness as a low scatter mass proxy. Based on the optical richness and the calibrated weak lensing mass--richness relation, mass of individual objects down to $\sim 10^{13}M_\odot$ can be estimated with a precision of $\sim 20$ per cent.
\end{abstract}


\section{Introduction}
\label{sec_intro}

Studies of number counts of galaxy clusters can put significant constraints on cosmological parameters \citep{vik+al09b,man+al15,xxl_XXV_pac+al18,cos+al19}. Detection methods can recover large numbers of galaxy clusters with high levels of purity and completeness from optical bands \citep{ryk+al14,ogu+al18,mat+al19}, X-ray data \citep{xxl_I_pie+al16}, or observations of the Sunyaev-Zel'dovich effect \citep{ble+al15,planck_2015_XXVII}. The constraining power of cluster abundance is strongly enhanced if the mass calibration is well understood and the mass--observable relation well known \citep{sar+al16}. An accurate and precise cosmological analysis requires the knowledge of both scaling parameters and intrinsic scatter, but recent efforts exploiting data from large surveys have been inconclusive \citep{planck_2015_XXIV,des_abb+al20}, the suspect culprit being a biased knowledge of the observable--mass scaling relation. 

The practical difficulties confront with a solid theoretical understanding of the main processes behind the scaling relations. In the self-similar scenario of structure virialisation  \citep{kai86,gio+al13,ett15}, the mass is the driving property which informs every other halo property. Tight scaling relations in form of power laws relate the cluster properties but non gravitational processes or deviation from equilibrium can affect the scalings and introduce an intrinsic scatter, such that, e.g. clusters with the same mass can have somewhat different optical richness or X-ray luminosity. Numerical simulations \citep{sta+al10,fab+al11,ang+al12,tru+al18} and observations \citep{mau14,man+al16,ser+al20_hscxxl} show that the intrinsic scatter is approximately log-normal.

One of the main problems in cosmological analyses is the selection of a complete sample of galaxy clusters with well measured masses. Weak lensing (WL) masses are regarded as reliable measurements but analyses of individual halos are challenging due to low signal-to-noise detections. Accurate and precise results can be available only for either heterogeneous or small samples of massive objects \citep{wtg_III_14,ume+al14,ok+sm16,mel+al17,ser+al17_psz2lens}. Even with high quality data, the analysis of small groups suffers from very large statistical uncertainties \citep{ume+al20,ser+al20_hscxxl}. Furthermore, projection effects, triaxiality, and prominent substructures can bias the mass measurement \citep{ras+al12}.

The signal of different galaxy clusters at fixed observables can be coherently added. This stacking technique allows to significantly enhance the signal of less massive halos and to measure their average mass in a range that is out of reach for analyses of individual clusters \citep{ma+se07,joh+al07,roz+al11,mel+al17,sim+al17a}. Results are very solid. Stacked WL masses and optical richnesses show tight correlations \citep{mcc+al19,mur+al19,bel+al19}. However, the picture is still not complete, since most stacking analyses do not return the estimate of the intrinsic scatter, which is usually assumed to be negligible. Whereas the intrinsic scatter of stacked quantities can be negligible, the scatter of individual halos is not and we need to know it for cosmological inference based on number counts.


Here, we tackle the problem of how to reconstruct the full information on scaling relations and intrinsic scatter from stacked data. As a study case, we apply the method to the galaxy clusters detected with the algorithm AMICO \citep[Adaptive Matched Identifier of Clustered Objects,][]{bel+al18,mat+al19} in KiDS Data Release 3 \citep[Kilo-Degree Survey,][]{dej+al13,kui+al15}, a WL survey in the Southern hemisphere. This is the sixth in the CoMaLit (COmparing MAsses in LITerature) series of papers, wherein we have been applying Bayesian hierarchical procedures to studies of masses and scaling relations. The method can deal with heteroscedastic and possibly correlated measurement errors, intrinsic scatter, upper and lower limits, systematic errors, missing data, forecasting, time evolution, and selection effects. In the first paper of the series \citep[ CoMaLit-I]{se+et15_comalit_I}, we considered the calibration of scaling relations and we assessed the level of intrinsic scatters in WL or X-ray mass proxies. In the second paper of the series \citep[ CoMaLit-II]{ser+al15_comalit_II}, we introduced the Bayesian method to infer scaling relations and we applied it to WL clusters with measured SZ flux. The third paper of the series \citep[ CoMaLit-III]{ser15_comalit_III} presented the Literature Catalogs of weak Lensing Clusters of galaxies (LC$^2$), a meta-catalog of WL clusters. The fourth paper of the series \citep[ CoMaLit-IV]{se+et15_comalit_IV} dealt with redshift evolution and completeness. The scalings of optical richness, X-ray luminosity, and galaxy velocity dispersion with mass were considered. In the fifth paper of the series \citep[ CoMaLit-V]{se+et17_comalit_V} we dealt with efficient mass forecasting. The method was extended to multi-dimensional analyses in \citet{xxl_XXXVIII_ser+al19}.

The paper is as follows. In Sec.~\ref{sec_scat}, we discuss proxies, intrinsic scatters, and degeneracies which affect parameter recovery. The stacking technique is introduced in the framework of a Bayesian model in Sec.~\ref{sec_stac}. The method to recover the intrinsic scaling relation from stacked data is presented in Sec.~\ref{sec_reco}. In Sec.~\ref{sec_amico_kids}, we consider the optically detected clusters in the AMICO-KiDS-DR3 catalog \citep{mat+al19}. In Sec.~\ref{sec_prox}, we review some results from literature. Section~\ref{sec_conc} is devoted to some final considerations. In App.~\ref{sec_biva}, we present alternative expressions for the bivariate normal distribution of two scattered proxies. In App.~\ref{sec_syst}, we detail how systematics uncertainties are dealt with in the CoMaLit approach. In App.~\ref{sec_repr}, we provide information to reproduce the paper results.

\subsection{Notation and conventions}

As reference cosmological model, we assume a flat $\Lambda$CDM ($\Lambda$ and Cold Dark Matter) universe with matter density parameter $\Omega_\text{M}=0.3$, and Hubble constant $H_0=70~\text{km~s}^{-1}\text{Mpc}^{-1}$.


The notation `$\log$' represents the logarithm to base 10 and `$\ln$' is the natural logarithm. Scatters in natural logarithm can be quoted as percents. Throughout the paper, unless otherwise noted, we denote $\sigma$ as the intrinsic scatter in $\log$ (decimal) quantities and use $\delta$ to represent $\log$ (decimal) measurement uncertainty. 

Unless stated otherwise, central values and dispersions of the parameter distributions are computed using the bi-weighted statistics \citep{bee+al90} of the marginalised posterior distributions.

Computations were performed with the \textsc{R}-package \texttt{LIRA}.\footnote{The package \texttt{LIRA} (LInear Regression in Astronomy) is publicly available from the Comprehensive R Archive Network at \url{https://cran.r-project.org/web/packages/lira/index.html}. For further details, see \citet{ser16_lira}.} As baseline, we consider the standard priors used throughout the CoMaLit series, see e.g. \citetalias{se+et15_comalit_IV}.

\section{Proxies and intrinsic scatter}
\label{sec_scat}

\begin{figure}
       \resizebox{\hsize}{!}{\includegraphics{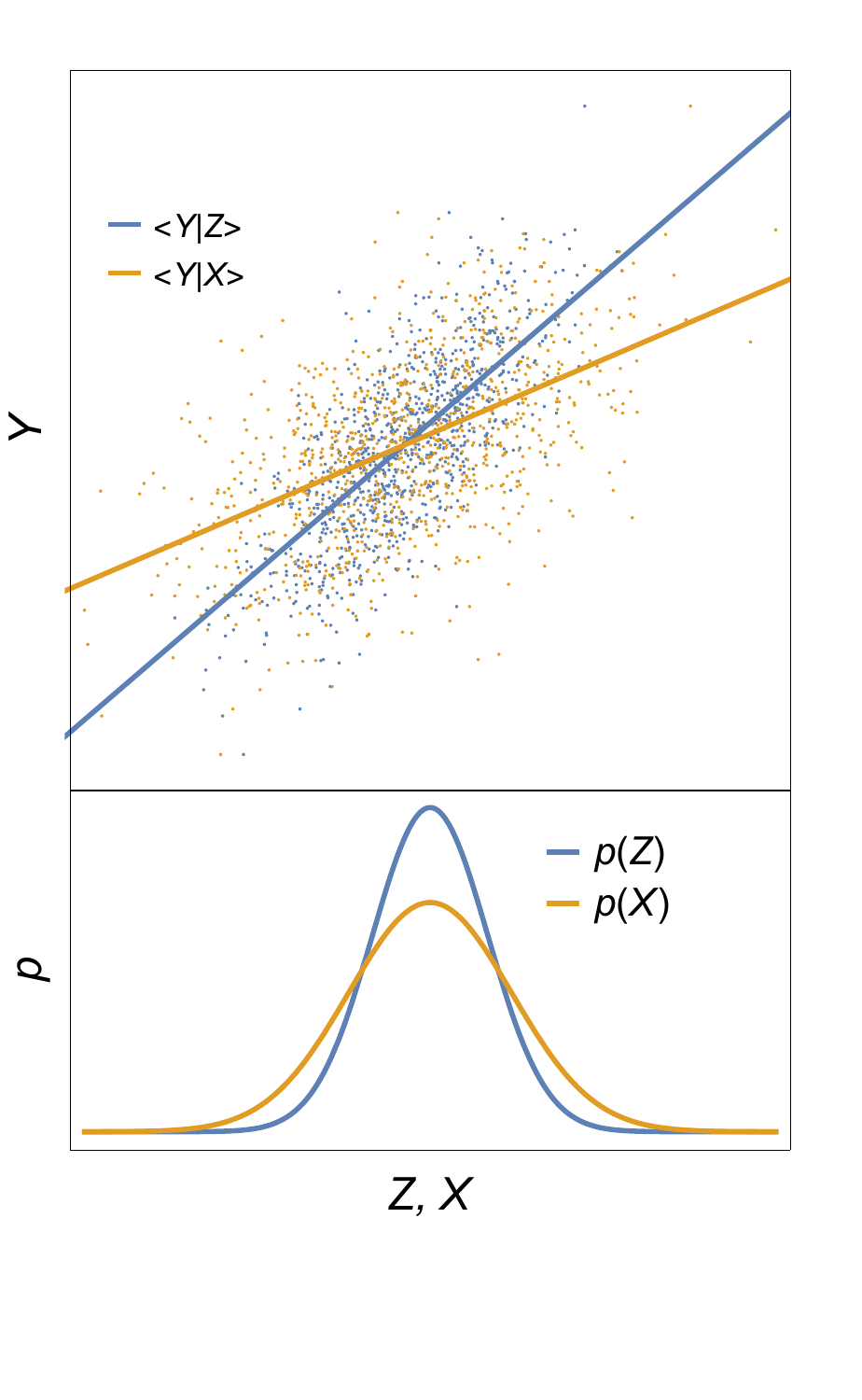}}
       \caption{Conditional scaling relations for scattered proxies in arbitrary units. $Z$ (turquoise) is the latent variable, which $Y$ and $X$ (orange) are scattered proxies of. Due to scatter, the distribution of $X$ is more extended than $Z$ (lower panel) and the scaling relation tracing the mean probability of $Y$ given $X$, $<Y|X>$ is flatter than $<Y|Z>$ (upper panel).}
	\label{fig_YIZ_YIX}
\end{figure}

\begin{figure}
       \resizebox{\hsize}{!}{\includegraphics{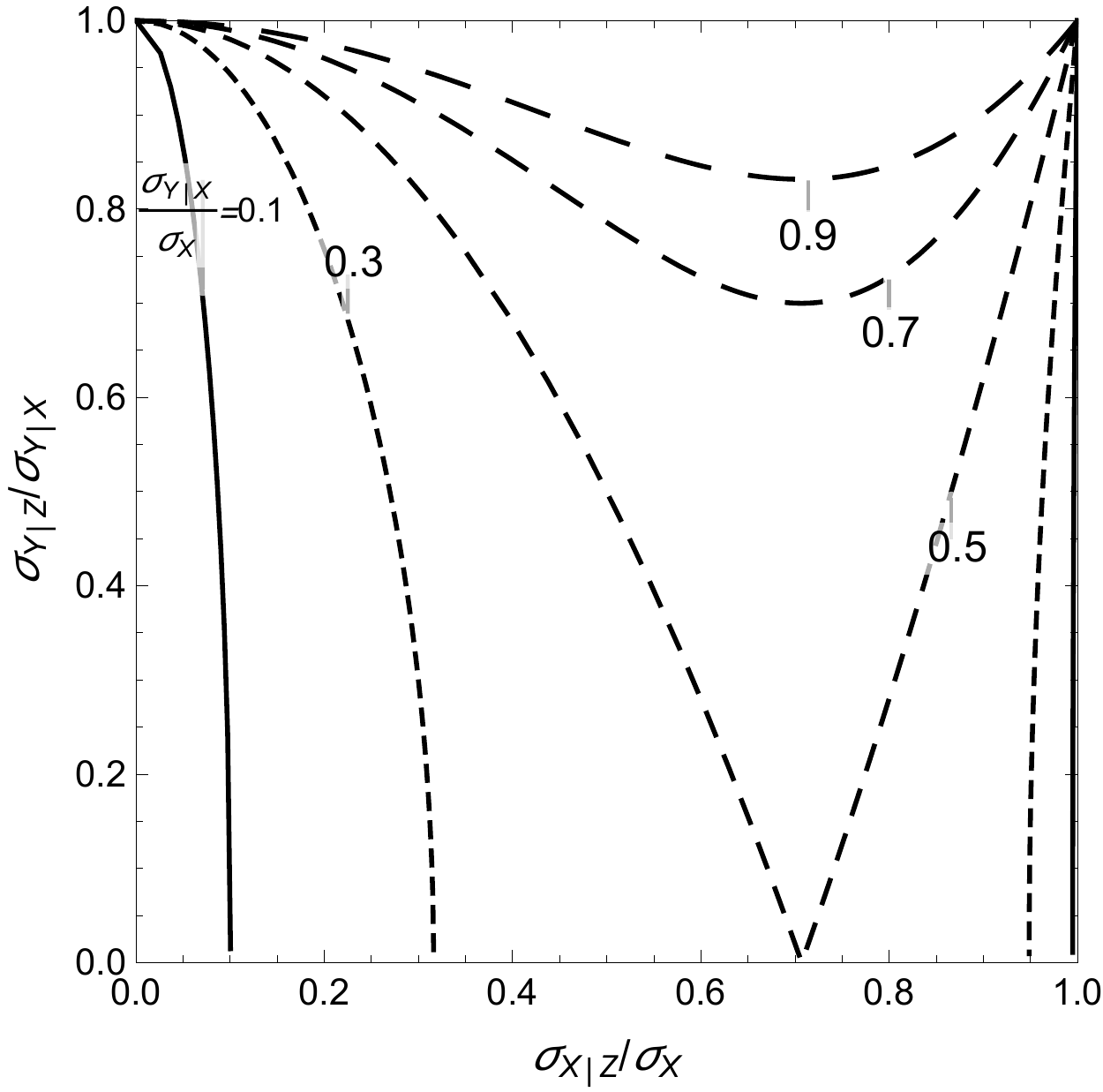}}
       \caption{Degeneracy problem in measuring the intrinsic scatters of two variables $X$ and $Y$, with respect to a third, hidden variable $Z$. The parametric plot shows $\sigma_{X|Z}$ (in units of the observable $\sigma_X$) and $\sigma_{Y|Z}$ (in units of the observable $\sigma_{Y|X}$) for $\sigma_Z$ spanning the range from 0 to $\sigma_X$. The contours are for values of $\sigma_{Y|X}/\sigma_X$ from 0.1 to 0.9 in steps of 0.2. Here we are considering $\beta_{Y|Z}=\beta_{X|Z}=1$. Smaller values of $\sigma_{X|Z}/\sigma_X$ correspond to larger values of $\sigma_Z/\sigma_X=(1-\sigma_{X|Z}^2/\sigma_X^2)^{1/2}$.}
	\label{fig_sigmaXIZ_sigmaYIZ}
\end{figure}

In most astrophysical analyses, we have to deal with scattered proxies of an underlying property. Here, we reconsider what already discussed in, e.g., \citet{edd13,mal22,jef38,edd40,ak+be96,kel07,an+be12,se+et15_comalit_I} and references therein. In this section, we neglect measurement uncertainties for simplicity.

Let us consider linear relations. We denote the intrinsic property as $Z$ and its scattered proxy as $X$. For a given $Z$, the expected value of $X$ is
\beq
\label{eq_cond_1}
<X|Z> = \alpha_{X|Z}+  \beta_{X|Z} Z,
\eeq
where $\alpha_{X|Z}$ and $\beta_{X|Z}$ are the normalisation and the slope of the $X$-$Z$ scaling relation, respectively. The intrinsic normal scatter is indicated as $\sigma_{X|Z}$. 

Let $Y$ be a second proxy related to $Z$ similarly to Eq.~(\ref{eq_cond_1}). Here we take for simplicity the intrinsic scatter of $Y$ given $Z$, $\sigma_{Y|Z}$, to be uncorrelated from $X$.

If the variable $Z$ is normally distributed with mean $\mu_Z$ and standard deviation $\sigma_Z$, the total probability distribution can be written as
\begin{align}
p(X,Y,Z) = & {\cal N}(Z |\mu_Z,\sigma_{Z}) {\cal N}(X | \alpha_{X|Z}+  \beta_{X|Z} Z, \sigma_{X|Z})  \label{eq_cond_2} \\
   \times & {\cal N}(Y | \alpha_{Y|Z}+  \beta_{Y|Z} Z, \sigma_{Y|Z} ), \nonumber
\end{align}
where ${\cal N}(x |\mu,\sigma)$ is the Gaussian distribution of the variable $x$ with mean $\mu$ and variance $\sigma^2$. In this basic picture, the distribution of $Z$ depends on the selection criteria and how we assembled the sample. On the other hand, the scalings between $Y$--$Z$ or $X$--$Z$ express the physical relationships between the cluster observables and can be seen as intrinsic. We expect to see the same scaling between $Y$ and $Z$ independently on how $Z$ was chosen. This picture is simplified since the scaling parameters and the scatter affecting the relation $Y$--$Z$ might depend on $Z$, and, consequently, on the selection criteria. Scaling parameters may depend on redshift and mass. For example, the relation between the gas and the total mass in galaxy clusters is steeper at the low mass end and the scatter is usually smaller for relaxed clusters \citep{man+al16,lov+al20,ser+al20_hscxxl}.

In a standard observational set-up, we do not have direct access to $Z$ (e.g. the true mass), but we can measure $X$ (e.g. the optical richness) and $Y$ (e.g. the WL mass), which are distributed as a bivariate Gaussian, see App.~\ref{sec_biva},
\beq
\label{eq_cond_3}
p(X,Y) = {\cal N}^\mathrm{(2)} (\left\{ X, Y \right\} | \left\{ \mu_X, \mu_Y \right\}, \bmath{\Sigma}_{XY} ),
\eeq
where ${\cal N}^\mathrm{(2)}$ is the bivariate Gaussian distribution and $\bmath{\Sigma}_{XY}$ is the scatter covariance matrix. The degree of correlation between $X$ and $Y$, $\rho_{XY}$, depends on how much the distribution in $Z$ is spread with respect to the intrinsic scatters. The broader the distribution, the less (relatively) important the effect of the intrinsic scatters, and the more correlated $X$ and $Y$ are.  The distribution of the pair $\left\{ Y, Z \right\}$ is compared to the distribution of $\left\{ Y, X \right\}$ in Fig.~\ref{fig_YIZ_YIX} for the case of sizeable intrinsic scatters $\sigma_{X|Z}$ and $\sigma_{Y|Z}$ with respect to dispersion of the $Z$ distribution, $\sigma_Z$.

Alternatively, the probability of $X$ and $Y$ can be expressed in terms of the conditional probability of $Y$ given $X$, see Fig.~\ref{fig_YIZ_YIX}. This is practical when we forecast the unknown value of $Y$ for a given known value of $X$. As showed in App.~\ref{sec_biva},
\beq
\label{eq_cond_4}
p(X,Y) ={\cal N}(Y | \alpha_{Y|X}+  \beta_{Y|X} X, \sigma_{Y|X}) {\cal N}(X |\mu_X,\sigma_{X}) .
\eeq

Whereas the relation between $X$ and $Z$ (or $Y$ and $Z$) depends on the involved physics only, the relation between $X$ and $Y$ is not universal and depends on the sample of $Z$ that we are considering through the parameters $\mu_Z$ and $\sigma_Z$.  The mean $\mu_Z$ affects the normalisation $\alpha_{Y|X}$; the scatter  $\sigma_Z$ affects the normalisation $\alpha_{Y|X}$, the slope $\beta_{Y|X}$, and the conditional scatter $\sigma_{Y|X}$, see Eqs.~(\ref{eq_Y|X_2}-\ref{eq_Y|X_4}).

\subsection{Parameter degeneracies}

Let us consider observations of galaxy clusters. The true mass can play the role of $Z$, as a latent variable we do not have direct access to. We can estimate the X-ray mass assuming equilibrium or the WL mass, and we can measure some other properties, e.g. the optical richness or the X-ray luminosity. These are all scattered proxies of the true mass and can play the role of $X$ or $Y$. 

The inversion problem of determining $P(X,Y,Z)$ from $P(X,Y)$ is severely under-constrained. For Gaussian distributions, we want to determine 8 parameters, i.e. $\alpha_{X|Z}$, $\beta_{X|Z}$, $\sigma_{X|Z}$, $\alpha_{Y|Z}$, $\beta_{Y|Z}$, $\sigma_{Y|Z}$, and $\mu_Z$, and $\sigma_{Z}$, see Eq.~\ref{eq_cond_2}, from the measurements of 5 observables, i.e. $\alpha_{Y|X}$, $\beta_{Y|X}$, $\sigma_{Y|X}$, $\mu_X$, and $\sigma_{X}$, see Eq.~\ref{eq_cond_4}.

The $Y$-$Z$ and $X$-$Z$ scalings cannot be unambiguously determined. Even in the very favourable case of negligible intrinsic scatter ($\sigma_{X|Z}\ll \sigma_Z$), we can only measure the ratio of the slopes with respect to $Z$ and a renormalised difference between the intercepts,
\begin{eqnarray}
\beta_{Y|X}& \simeq & \frac{\beta_{Y|Z}}{\beta_{X|Z}} \, , \\
\alpha_{Y|X} & \simeq & \alpha_{Y|Z}-  \alpha_{X|Z} \frac{\beta_{Y|Z}}{\beta_{X|Z}} \, .
\end{eqnarray}
In practical cases, we can often assume that one scaling, i.e. between $X$ and $Z$, is known. For example, if we are studying a randomly oriented sample of relaxed clusters, the WL mass ($X$) is an unbiased proxy of the true mass ($Z$), with $\beta_{X|Z}=1$ and $\alpha_{X|Z}=0$. Under this condition, $\mu_X=\mu_Z$. Only the intrinsic scatter $\sigma_{X|Z}$ has still to be measured. In the following, we will assume that  $\beta_{X|Z}=1$ and $\alpha_{X|Z}=0$ if not otherwise stated.

Even if the scaling between $X$ and $Z$ is fixed, residual degeneracies still hamper the inversion problem. When the conditional intrinsic scatter is sizeable with respect to the dispersion in $Z$ ($\sigma_{X|Z}\sim \sigma_Z$), as can be the case when we study scaling relations for the optical properties of clusters selected in a narrow mass range, the $Y$-$X$ relation can be significantly flatter than the $Y$-$Z$ relation, see Fig.~\ref{fig_YIZ_YIX}. This is a result of the magnitude-dependent Malmquist bias \citep{edd13,mal22,jef38,edd40,but+al05,se+et17_comalit_V}. Due to the scatter between $X$ and $Z$, the marginalised distribution in $X$, $P(X)$, has a larger dispersion than $P(Z)$, 
\beq
\sigma_X^2= \sigma_Z^2+ \sigma_{X|Z}^2,
\eeq
which flattens the slope of the $Y$-$X$ relation. The larger the scatter $\sigma_{X|Z}$, the flatter the $Y$-$X$ relation, 
\beq
 \beta_{Y|X}=  \beta_{Y|Z} \left( 1-  \frac{ \sigma_{X|Z}^2}{ \sigma_X^2} \right). 
\eeq
A degeneracy between the slope $\beta_{Y|Z}$ and the scatter $\sigma_{X|Z}$ then persists.

When we consider samples on a more extended range, e.g. spanning from the group scale to the more massive halos, the $Y$-$X$ relation is steeper with a larger absolute value of $\beta_{Y|X}$ than for limited samples in narrow $Z$ ranges. According to our simplified picture, the relation $Y$-$Z$ is universal, whereas the relation $Y$-$X$ depends on the properties of the sample.


There can be also a remaining degeneracy between the slope $\beta_{Y|Z}$ and the normalization $\alpha_{Y|Z}$,
\beq
\alpha_{Y|X}= \alpha_{Y|Z} + (  \beta_{Y|Z} - \beta_{Y|X} )\mu_X
\eeq
This degeneracy can be reduced with convenient unit of measurements for $X$ such that  $\mu_X \sim 0$.

We may want to study the scatter and the properties of the measurable quantities $X$ and $Y$ with respect to the latent $Z$ (e.g. the true mass, which is hidden to observations). This problem can be tackled with suitable priors or assumptions but, if we cannot directly measure $Z$ and $\sigma_Z$, the determination of the intrinsic scatters, $\sigma_{Y|Z}$ and $\sigma_{X|Z}$, is under-constrained, see Fig.~\ref{fig_sigmaXIZ_sigmaYIZ}, where we are considering the simplified case with $\beta_{Y|Z}=\beta_{X|Z}=1$. From the analysis of the distribution of $X$ and $Y$, we can constrain two independent standard deviations, e.g. $\sigma_{Y|X}$ and $\sigma_{X}$, but not the all three quantities that we are interested in, i.e. $\sigma_Z$, $\sigma_{Y|Z}$, and $\sigma_{X|Z}$. 

The value of the ratio $\sigma_{Y|X}/\sigma_{X}$ can be determined by observations, and the locus of possible solutions for the intrinsic scatters is then determined, i.e. one of the lines in Fig.~\ref{fig_sigmaXIZ_sigmaYIZ}, each one corresponding to a given value of the observable ratio. Each pair of (renormalised) values of  $\sigma_{Y|Z}$ and $\sigma_{X|Z}$ which lie along the degeneracy locus is compatible with observations.  Observations can determine the line where the intrinsic scatters lie in the parameter space shown in Fig.~\ref{fig_sigmaXIZ_sigmaYIZ} (e.g. either the full, or the dashed, or the long-dashed ones), but we cannot break the degeneracy along the line. The solution is even more complicated if the scatters in $X$ and $Y$ are correlated.

If the variable $X$ is not scattered ($\sigma_{X|Z}= 0$), then $\sigma_Z=\sigma_X$, and the scatter $\sigma_{Y|X}$ is equal to $\sigma_{Y|Z}$ (top left corner in Fig.~\ref{fig_sigmaXIZ_sigmaYIZ}). This corresponds to the case of $X$ being an unscattered proxy of $Z$. If the $Z$ distribution collapses to the Dirac delta function ($\sigma_Z\rightarrow 0$), then the scatter $\sigma_X$ equals the conditional scatter $\sigma_{X|Z}$, and $\sigma_Y=\sigma_{Y|Z}$, or equivalently, $\sigma_{Y|Z}=\sigma_{Y|X}$ (top right corner in Fig.~\ref{fig_sigmaXIZ_sigmaYIZ}). 

The degeneracy can be partially broken by an optimised set-up. If we are studying a scaling relation, we are likely studying a convenient sample where we are confident (based e.g. on external information) that the intrinsic scatter is smaller than the dispersion of the sample ($\sigma_{X|Z}< \sigma_X$) and that the intrinsic scatter of $Y$ given $X$ ($\sigma_{Y|X}$) is of the same order as, even though a bit smaller than, $\sigma_{Y|Z}$, the intrinsic property we want to infer. In practice, we have to confine ourself to the upper left corner of the parametric space shown in Fig.~\ref{fig_sigmaXIZ_sigmaYIZ} to better constrain the intrinsic scatters.

Suitable priors can also limit the parameter degeneracies. This can be the case for non-informative priors too. Priors for positive defined quantities which are nearly constant in the log space, i.e. $P_\text{prior}(\log \sigma_{X|Z}) \sim$ constant over an extended parameter domain, are regarded as non informative since they allow for very large or small scatters. Nevertheless, these priors slightly favour smaller values of $\sigma_{Y|Z}$, $P_\text{prior}( \sigma_{X|Z}) \sim 1/\sigma_{X|Z}$, and, e.g., the left side of Fig.~\ref{fig_sigmaXIZ_sigmaYIZ}. As prior for the variances, we adopt an inverse Gamma distribution \citepalias{ser+al15_comalit_II}.

\section{Stacking}
\label{sec_stac}

\begin{figure}
	\begin{tabular}{c}
       \resizebox{\hsize}{!}{\includegraphics{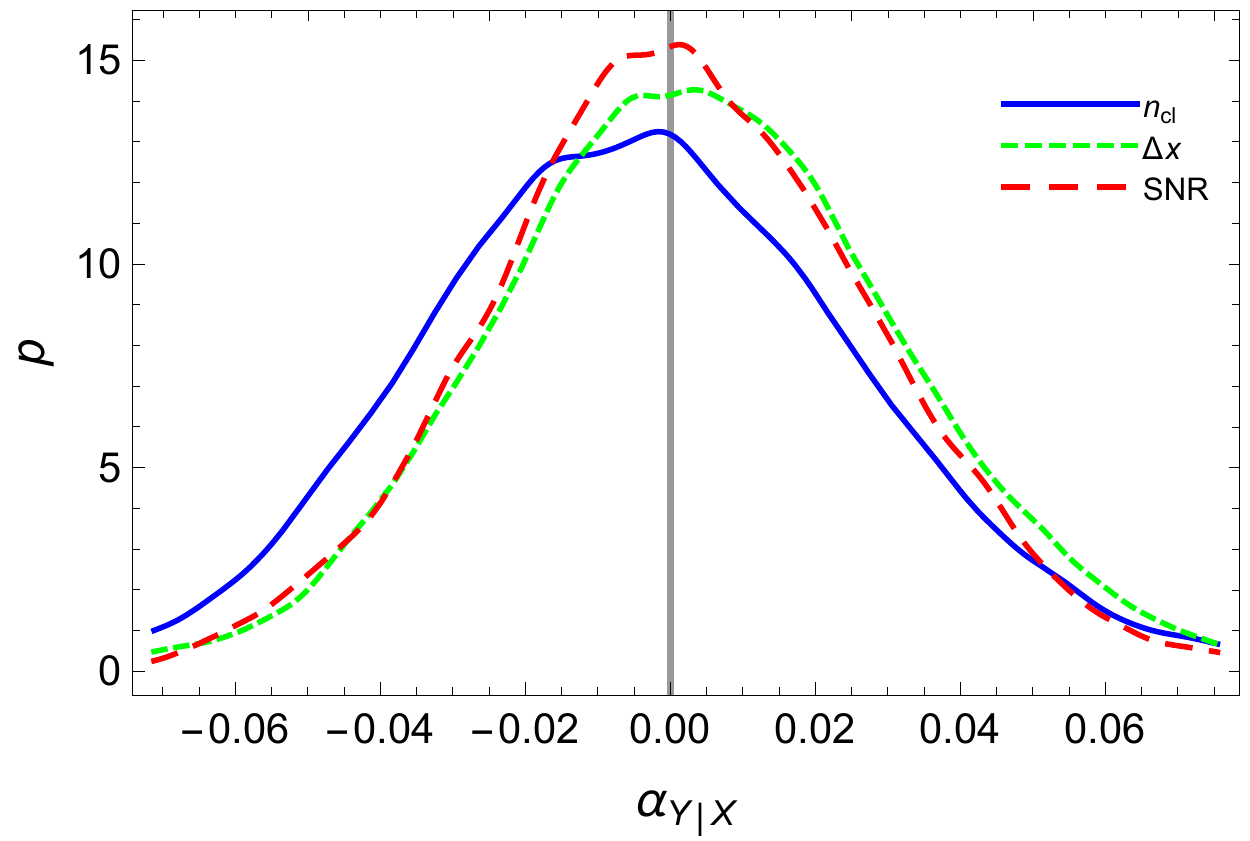}} \\
       \resizebox{\hsize}{!}{\includegraphics{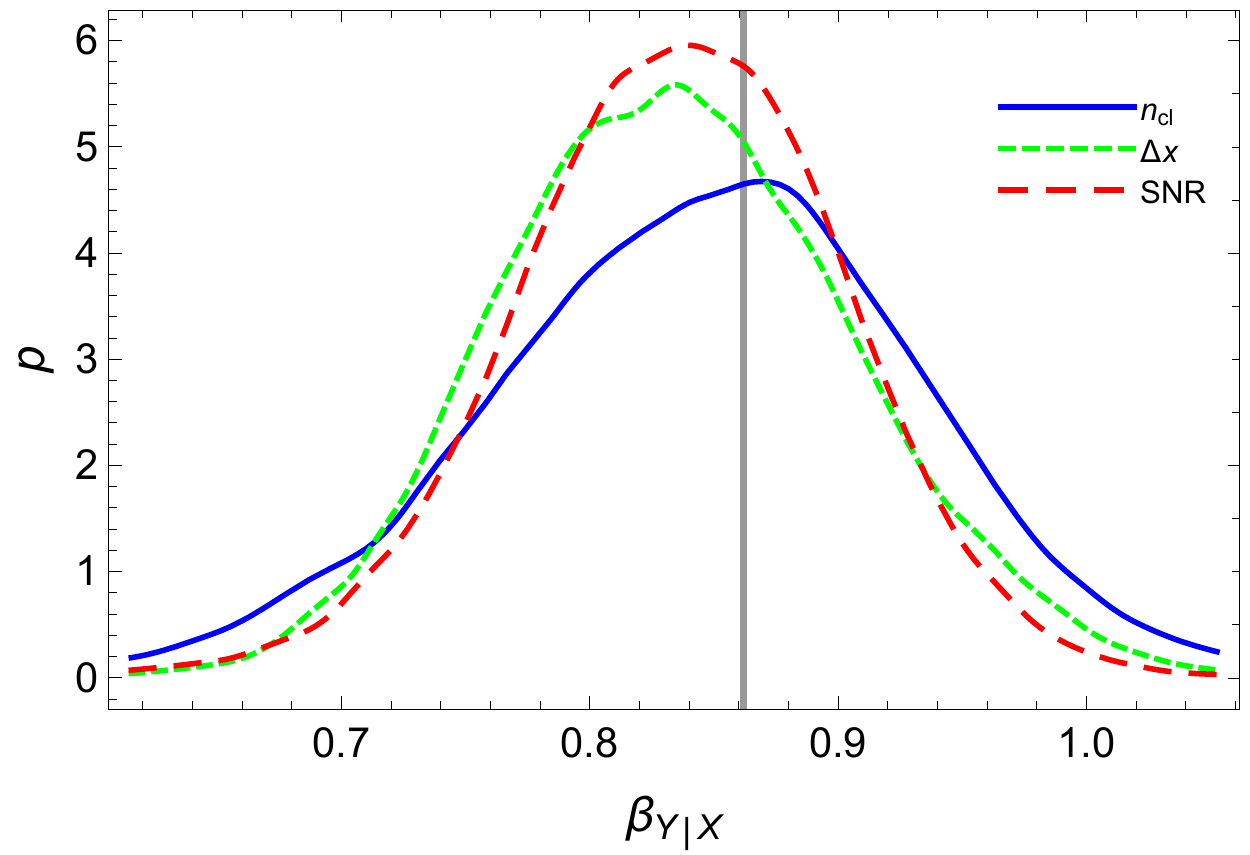}} \\
        \resizebox{\hsize}{!}{\includegraphics{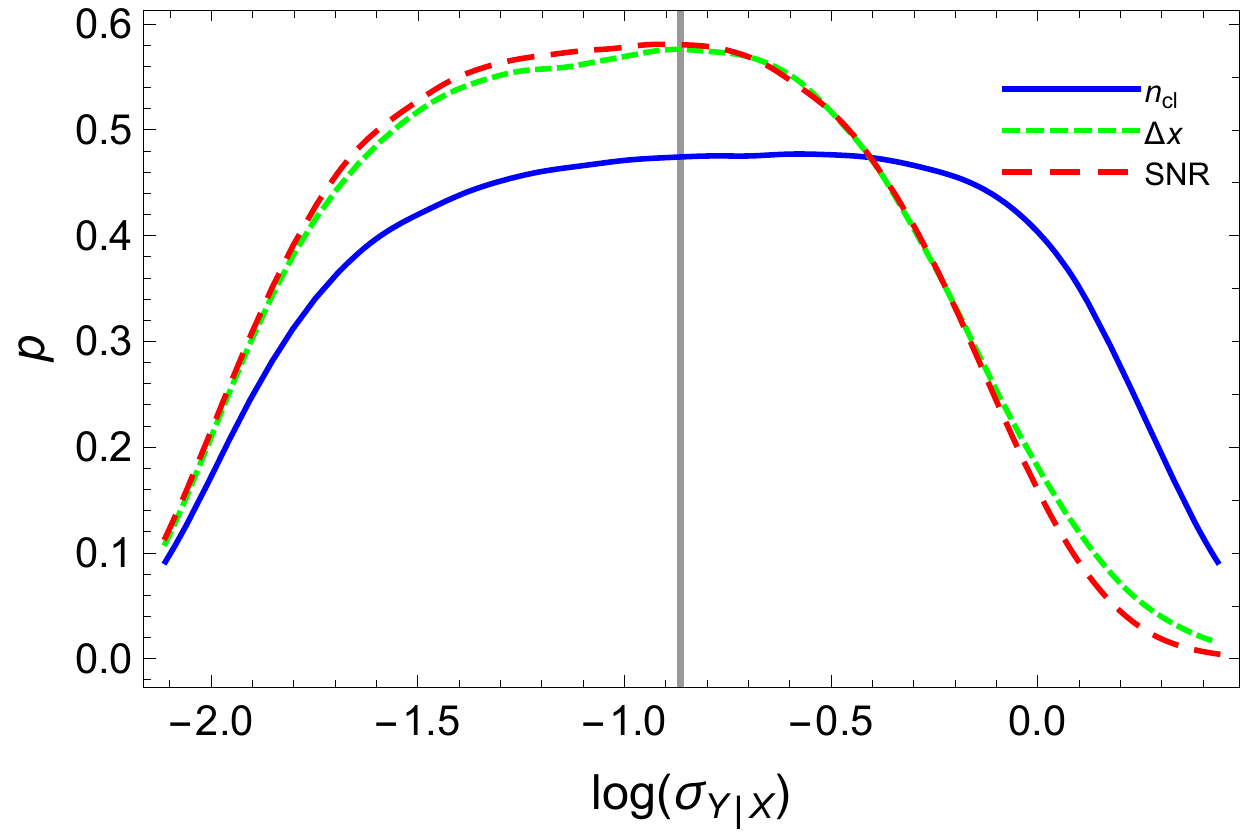}} 
        \end{tabular}
        \caption{Probability density functions of the parameters of the conditional scaling relation $Y|X$ recovered from a sample of 10 simulated stacked data $\{\bar{X}_i, \bar{Y}_i \}$. Results are averaged over $10^2$ mocks. The full blue line, the short dashed green line, and the long dashed red line refer to a binning scheme which is either uniform for number of clusters per bin, or interval length, or bin SNR, respectively. The vertical grey lines mark the input parameters. From the top to the bottom panel, we plot the a posteriori marginalised probability of normalisation $\alpha_{Y|X}$, slope $\beta_{Y|X}$, and scatter $\sigma_{Y|X}$.}
	\label{fig_PDF_stacking}
\end{figure}

For most samples of galaxy clusters, we are provided with reliable measurements of some properties, e.g. position, redshift, luminosity, but other properties, e.g., the WL mass, cannot be accurately measured for individual clusters. We then add the signal of the clusters which are similar with regard to one measured property to estimate the mean value of the property we cannot individually measure.  For example, we can measure the optical richness of individual optically selected clusters, but not the WL signal that has to be recovered from the stacked signal of a subsample of objects with similar values of richness.

Let $X$ be the proxy we can individually measure and $Y$ the proxy we want to constrain. We stack the signal produced by all the clusters with $X_{\text{min},i}< X \le X_{\text{max},i}$, where $X_{\text{min},i}$ and $X_{\text{max},i}$ are the lower and upper boundaries of the $i$-th bin, respectively. We end up with a sample of known mean values, $\{\bar{X}_i, \bar{Y}_i \}$. Assuming that all clusters are weighted only by their number:
\begin{eqnarray}
\bar{X}_i & =& \frac{ \int_{X_{\text{min},i}}^{X_{\text{max},i}} X  p(X) d X}{ \int_{X_{\text{min},i}}^{X_{\text{max},i}}  p(X) d X},  \label{eq_stack_1} \\
\bar{Y}_i & =& \frac{\int_{-\infty}^{+\infty} \int_{X_{\text{min},i}}^{X_{\text{max},i}} Y  p(X,Y) d X d Y}{ \int_{X_{\text{min},i}}^{X_{\text{max},i}}  p(X) d X} .   \label{eq_stack_2}
\end{eqnarray}
By definition, $\bar{Y}_i$ is the expected value for a given bin, i.e. for a given $\bar{X}_i$. For the normal distributions discussed in Sec.~\ref{sec_scat},
\begin{eqnarray}
\bar{X}_i & = & \mu_X \nonumber \\
& +&  \sigma_X^2 \frac{ {\cal N}(X_{\text{min},i} | \mu_X, \sigma_X ) -{\cal N}(X_{\text{max},i} | \mu_X, \sigma_X ) } {\frac{1}{2} \left[ \mathrm{erf}\left( \frac{X_{\text{max},i}-\mu_X}{\sqrt{\sigma_X}} \right) -  \mathrm{erf}\left( \frac{X_{\text{min},i}-\mu_X}{\sqrt{\sigma_X}} \right) \right] } \,  ,   \label{eq_stack_3} \nonumber \\
\bar{Y}_i & = &  \alpha_{Y|X}+  \beta_{Y|X} \bar{X}_i \,  .\label{eq_stack_4}
\end{eqnarray}

We assume that the intrinsic scatter $\sigma_{Y|X}$ is constant and uncorrelated. Then, $\bar{Y}_i$ is affected by an intrinsic scatter $\sigma_{\bar{Y}|\bar{X},i}=\sigma_{Y|X}/\sqrt{N_{\text{stack},i}}$, where $N_{\text{stack},i}$ is the number of clusters in the $i$-th bin.

The sample $\{ \bar{X}_i, \bar{Y}_i \}$ depends on the binning scheme, i.e. how we choose the boundaries of the intervals in $X$. The distribution $p(\bar{X}, \bar{Y})$ can differ from the unbinned $p(X,Y)$. They are equivalent if we choose the binning in such a way that $p(\bar{X})$ follows $p(X)$, and, as a consequence, the relations $\bar{Y}$-$\bar{X}$ and $Y$-$X$ are equivalent, i.e., for linearly related proxies, $\alpha_{\bar{Y}|\bar{X}}=\alpha_{Y|X}$ and $\beta_{\bar{Y}|\bar{X}}=\beta_{Y|X}$. This is the case if the bin boundaries are quantiles of $p(X)$.

Binning in quantiles can be unpractical if the signal-to-noise ratio (SNR) in the low value bins is too small for a precise measurement of $\bar{Y}$. If $X$ is the logarithm of some cluster property, e.g. the optical richness, and the signal is linear, i.e. it is proportional to $10^X$, the signal-to-noise ratio of the $i$-th bin can be written as
\beq
\text{SNR}_i  \propto \sqrt{N_{\text{stack},i}}   \langle 10^X \rangle_i ,
\eeq
where we have assumed that the noise per cluster is constant and uncorrelated. It can be convenient to stack the data in such a way that the signal-to-noise ratio is constant per bin in order to keep the relative uncertainty constant too.

To quantify how much the binning scheme hampers the recovery of the underlying $p(Y|X)$ distribution, we run $10^2$ simulations, each one with $N_\text{cl}=10^3$ clusters. Here, we distinguish the proxy $X$ from the result of the measurement process $x$, which differs from $X$ for the statistical uncertainty $\delta_X$. In analogy, we define $Y$ and $y$. The measurement results for the stacked quantities, $\bar{X}$ and  $\bar{Y}$ are $\bar{x}$ and $\bar{y}$, respectively. 

We set the underlying distribution of $Z$ as a normal distribution with $\mu_Z=0$ and $\sigma_Z=0.25$. The proxy $X$ is randomly distributed around $Z$ with $\alpha_{X|Z}=0$, $\beta_{X|Z}=1$, $\sigma_{X|Z}=0.1$. The observed $x$ were randomly distributed around $X$ assuming a constant statistical uncertainty of $\delta x=0.1/\ln(10)$. The second proxy $y$ is produced similarly to $x$. 

Finally, we assume that the statistical uncertainties on the stacked $\bar{y}_i$ are inversely proportional to the bin SNR, and $\delta\bar{y} =0.1/\ln(10)$ when $\text{SNR}  = S_\text{tot} / N_\text{bin} / \sqrt{N_\text{cl} / N_\text{bin}}$, where $ S_\text{tot}= \sum_{j=1}^{N_\text{cl}} 10^x_j$.

We stack the data in $N_\text{bin}=10$ bins defined according to the measured $x$. The bins are chosen such that: i) $p(\bar{x})$ follows $p(x)$, i.e. the bin boundaries are quantiles of $p(x)$; ii) the bins are equally spaced, i.e. $\Delta x_i= x_{\text{max},i}-x_{\text{min},i}=$ constant; iii) the SNR per bin is constant. 

Results are summarised in Fig.~\ref{fig_PDF_stacking}. The regression exploits a Bayesian inference method with non-informative priors, see App.~\ref{sec_repr}, where we consider the variable $Y$ as a scattered proxy of the measured $X$. Since we are interested in the conditional probability of $Y$ given $X$, we fit only two variables ($X$ and $Y$) and we neglect the latent $Z$. In the \texttt{LIRA} fitting we identify $X$ as $Z$, i.e. we put $\alpha_{X|Z}=0$, $\beta_{X|Z}=1$, and we neglect the scatter $\sigma_{X|Z}$, see App.~\ref{sec_repr}. The parameters of the scaling relation between $X$  and $Y$ are well recovered from the stacked data. This is expected for the properties of the stacked technique, for which $Y_i = \langle Y| X_i \rangle$ by design. Moreover, we find that the intrinsic scatter can be recovered for all binning schemes without any significant bias. The impact of the binning scheme is then negligible in most practical cases, and we can optimise the scheme to increase the statistical accuracy in the measurement of $\bar{Y}$.

\section{Reconstruction}
\label{sec_reco}

\begin{table}
\caption{Parameters of the scaling relation $Y$--$Z$ and $X$--$Z$ as recovered from simulated stacked data. Col. 1: parameter name. Col. 2: input value. Col. 3: results of fitting when the $y$ values are measured for individual objects. Col. 4: scaling parameters as derived from stacked data by fitting a population of fictitious $y$ produced with the method of Sec.~\ref{sec_reco}.}
\centering
	\begin{tabular}[c]{l l r@{$\,\pm\,$}l  r@{$\,\pm\,$}l}
	\hline
 \noalign{\smallskip}  
	 parameter & input	& \multicolumn{2}{c}{$y$ observed}	&	\multicolumn{2}{c}{$y$ recovered}  \\ 
	 \hline
	\noalign{\smallskip}  
$\alpha_{Y|Z}$	&	 [0]	&	0.00	&	0.02	&	0.00	&	0.04	\\
$\beta_{Y|Z}$	&	 [1]	&	1.00	& 	0.16	&	1.04	&	0.33 \\
$\sigma_{Y|Z}$	&	 [0.1]	&	0.09	&   0.05	&	0.13	& 0.11	\\
$\sigma_{X|Z}$	&	 [0.1]	&	0.09	&   0.05	&	0.11	& 0.07	\\
$\mu_{Z}$		&	 [0]	&	0.00	&   0.04	&	0.00	& 0.04	\\
$\sigma_{Z}$	&	 [0.25]	&	0.25	&   0.03	&	0.24	& 0.05 \\
\hline
\end{tabular}
\label{tab_summary_single}
\end{table}

\begin{figure*}
\centering
\resizebox{\hsize}{!} {
	\begin{tabular}{cc}
       \resizebox{.5\textwidth}{!}{\includegraphics{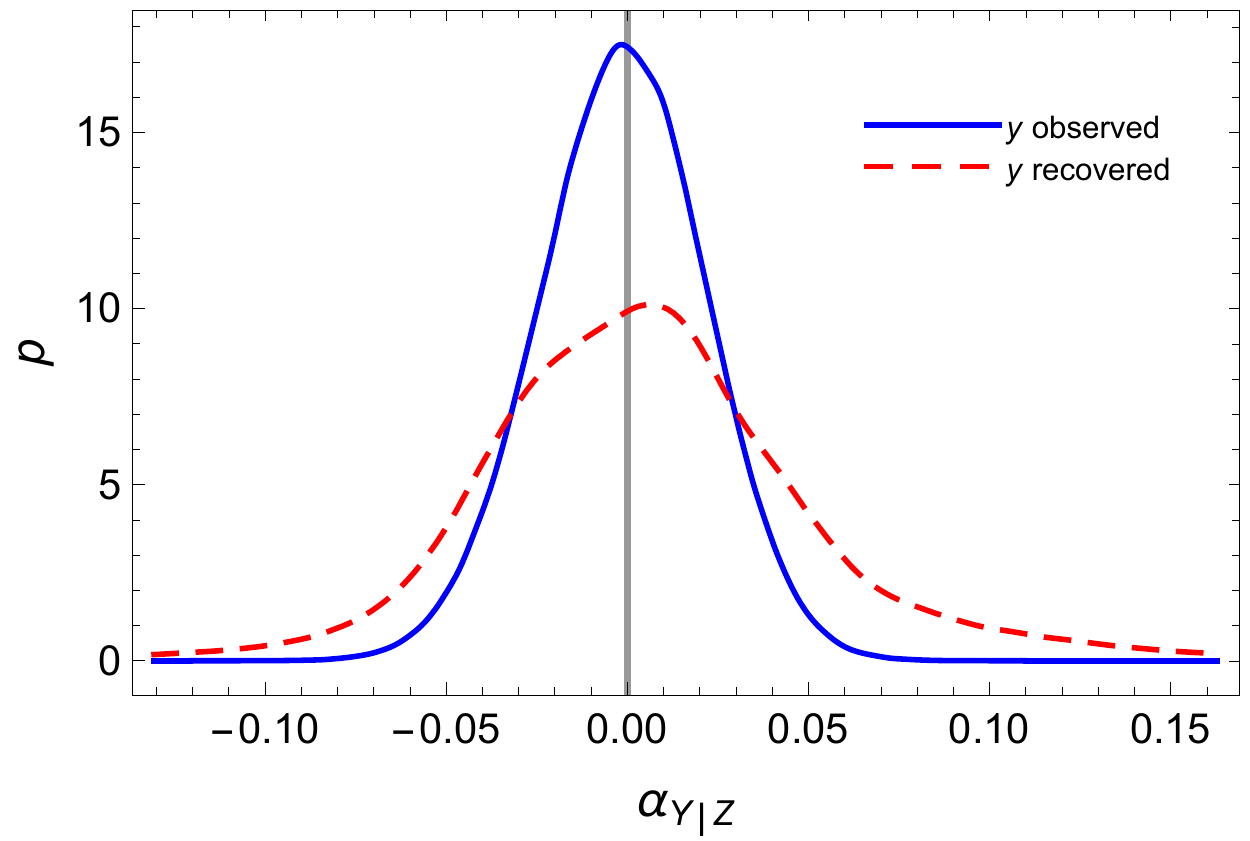}} &
       \resizebox{.5\textwidth}{!}{\includegraphics{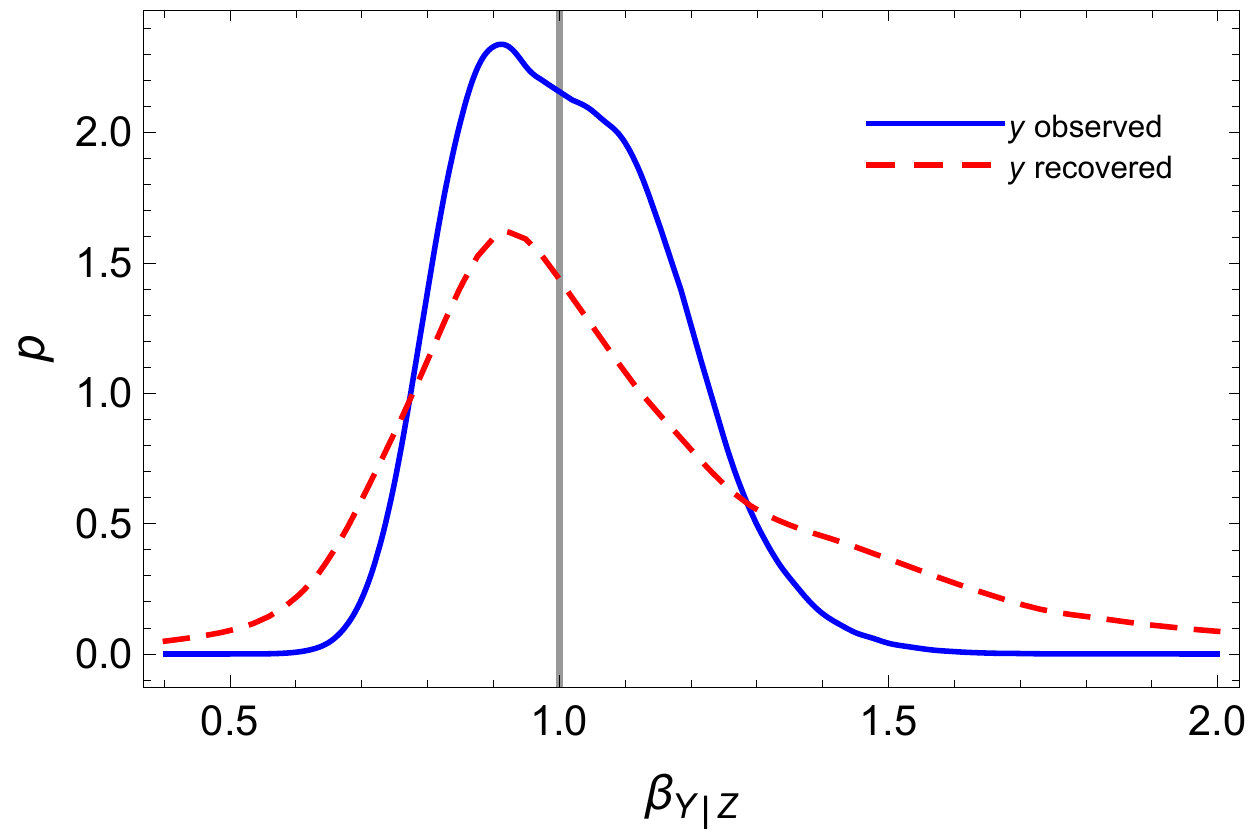}} \\
       \resizebox{.5\textwidth}{!}{\includegraphics{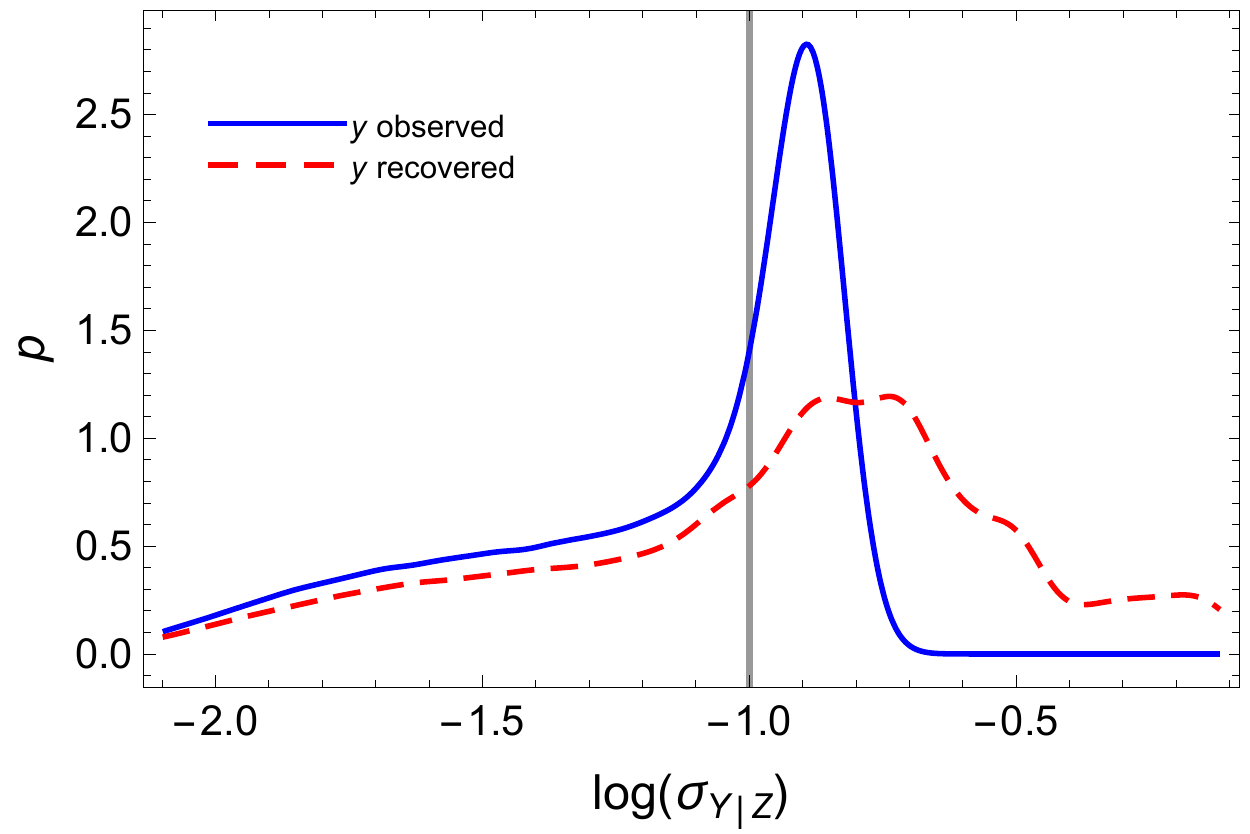}} &
       \resizebox{.5\textwidth}{!}{\includegraphics{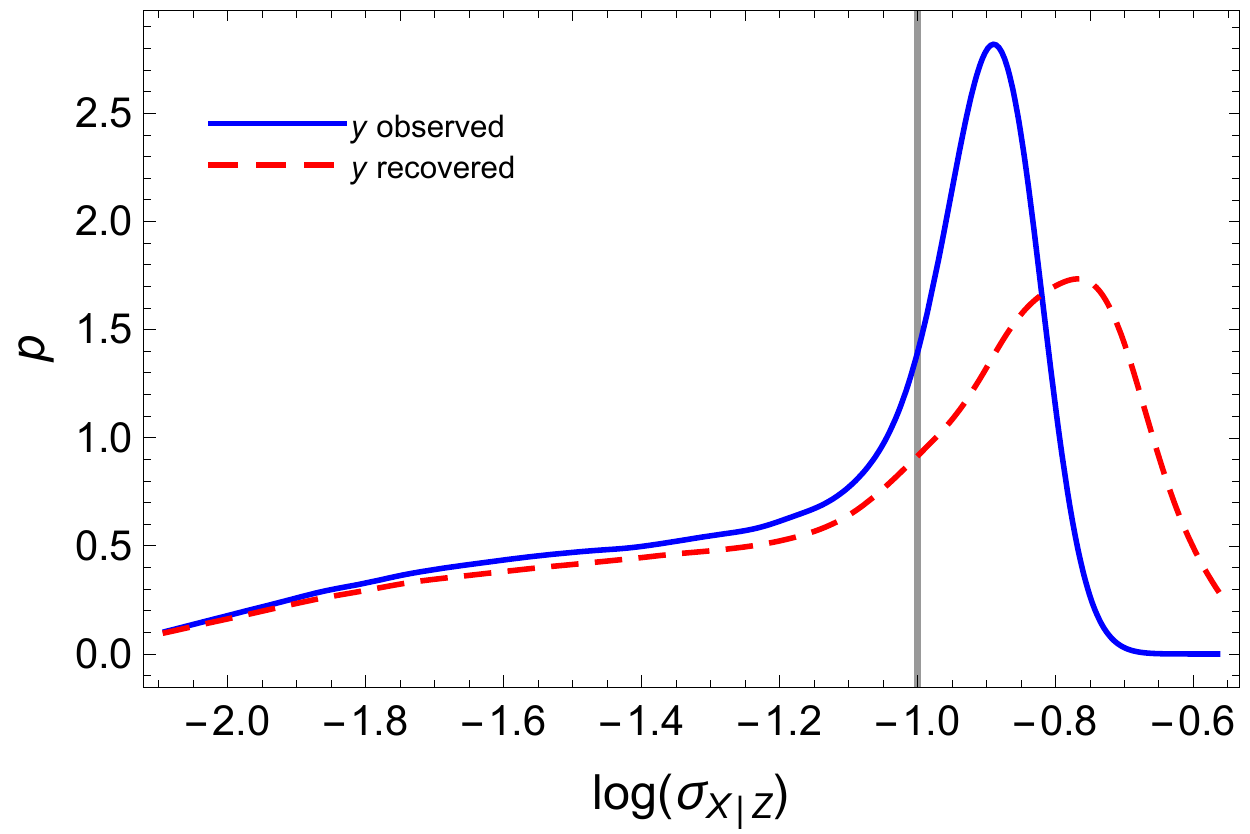}}  \\
        \resizebox{.5\textwidth}{!}{\includegraphics{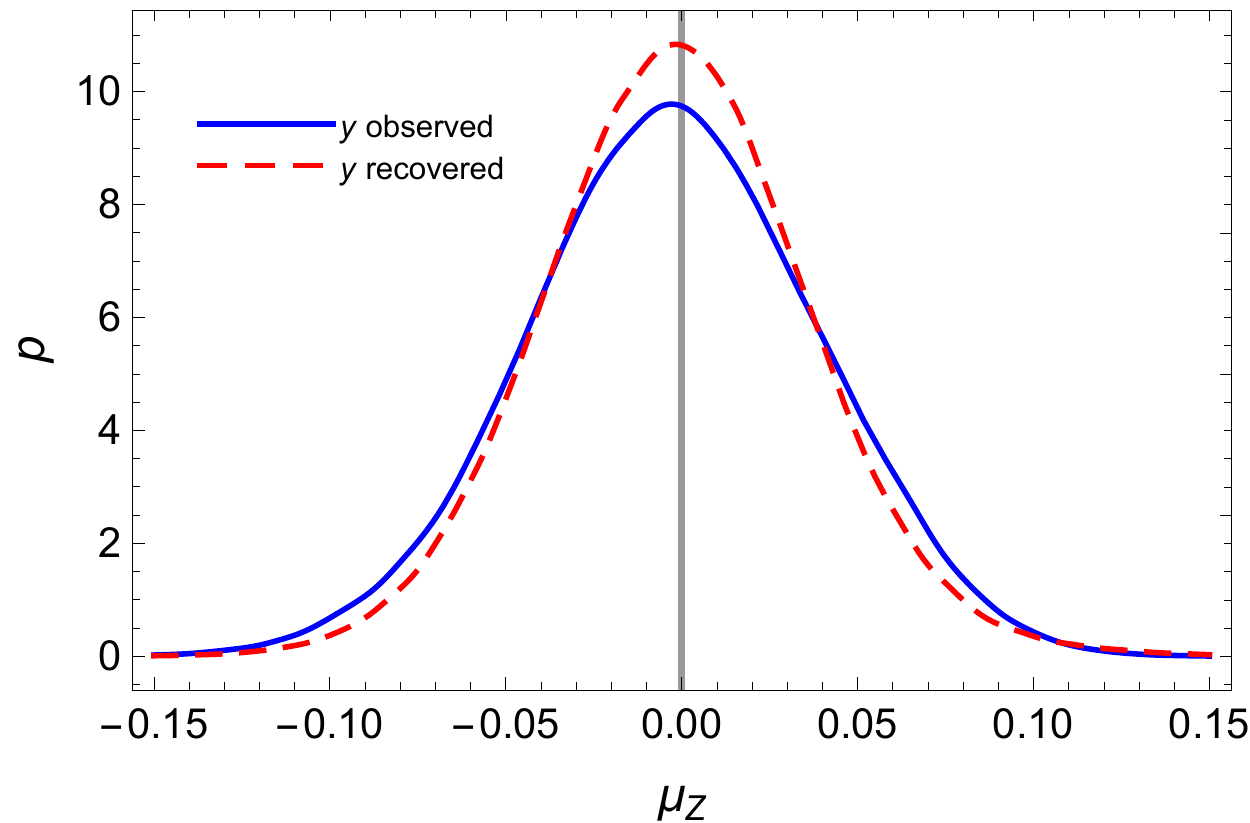}}  &
        \resizebox{.5\textwidth}{!}{\includegraphics{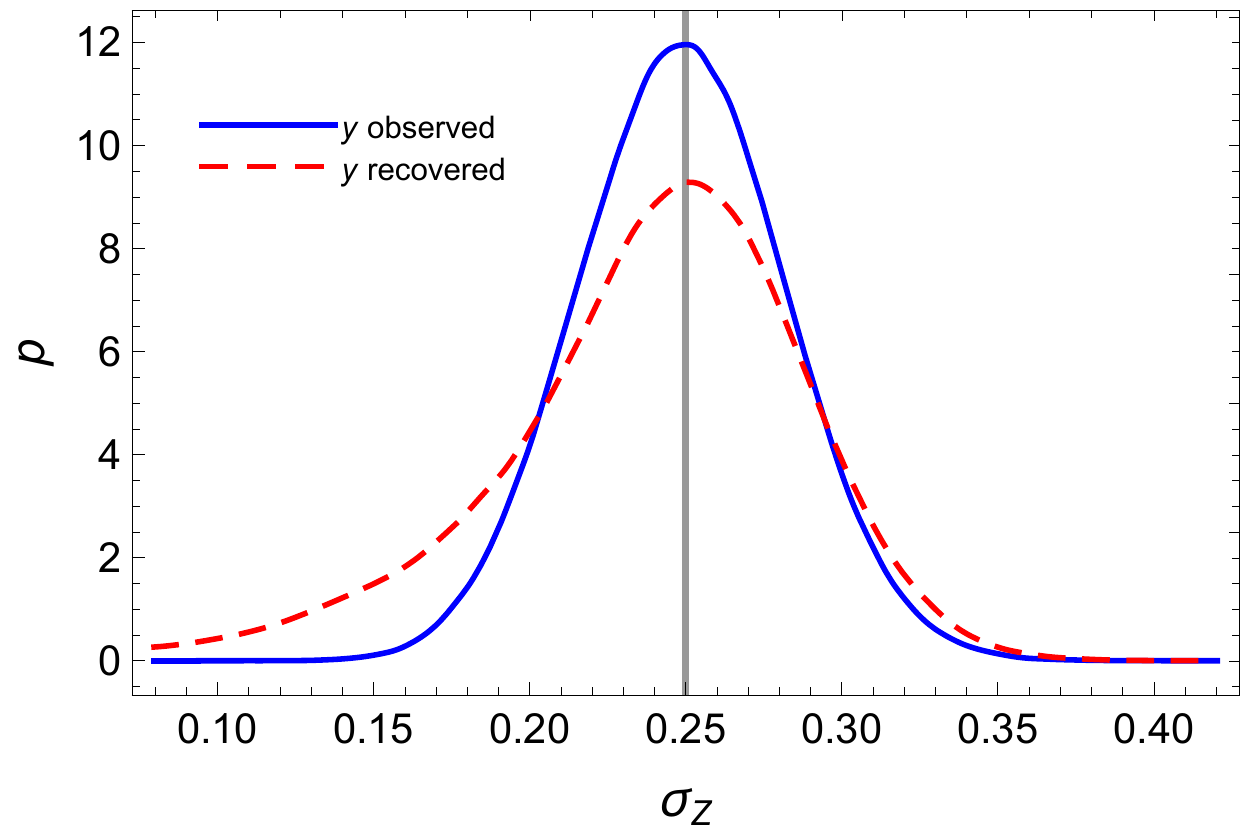}}  \\
        \end{tabular}
        }
        \caption{Probability density functions of the parameters of the scaling relation $Y$--$Z$ and $X$--$Z$ recovered from a simulated sample of 10 stacked data $\{\bar{X}_i, \bar{Y}_i \}$. Results are averaged over $10^2$ mocks. The full blue line, and the dashed red line refer to the fitting of $y$ values which are either observed for individual clusters ($y$ observed ) or reconstructed from the stacked data ($y$ recovered), respectively. The vertical grey lines mark the input parameters. We plot the a posteriori marginalised probability of the normalisation $\alpha_{Y|Z}$ (top left panel), of the slope $\beta_{Y|Z}$ (top right), of the intrinsic scatter $\sigma_{Y|Z}$ (middle left), of the intrinsic scatter $\sigma_{X|Z}$ (middle right), and of the mean $\mu_Z$ (bottom left) and standard deviation (bottom right) of the distribution $p(Z)$.}
	\label{tab_PDF_stacking}
\end{figure*}

In this section, we describe a three-steps procedure to recover the distribution $p(X,Y|Z)$ of two scattered proxies $X$ and $Y$ of an underlying property $Z$ based on stacked data, $\{\bar{x}, \bar{y}\}$, and the distribution of one individually measured proxy, $p(x)$.

We first fit the stacked data to recover the conditional $p(Y|X)$. This is done as described in Sec.~\ref{sec_stac}. As a result of the regression, we constrain the parameters of the scaling relation, $\alpha_{Y|X}$ and $\beta_{Y|X}$, and the scatter $\sigma_{Y|X}$. 

As a second step, we generate a fictitious population of $y$ based on the observed $x$ and on the conditional $p(Y|X)$ derived in the first step. Given each observed $x$, we draw a fictitious $Y_\text{f}$ thanks to $P(Y|X)$. Since we are using $x$ instead of $X$, we associate an uncertainty $\delta y_\text{f} =|\beta_{Y|X}| \delta x$. The correlation between $\delta x$ and $\delta y_\text{f}$ is $\beta_{Y|X}/|\beta_{Y|X}|$. If the results of the first step are in the form of a Monte-Carlo chain, each $y_\text{f}$ can be extracted by adopting a set of parameters of the $Y$-$X$ relation randomly drawn from the chain.

As a third and final step, we fit the observed $x$ and the fictitious $y_\text{f}$  to recover the relations of the proxies with the latent $Z$, i.e. the parameters which characterise $p(X,Y|Z)$.

To test the procedure, we run $10^2$ simulations, each one with $N_\text{cl}=10^3$ data points, with the same set-up described in Sec.~\ref{sec_stac}. For comparison, we also consider the case when the proxy $Y$ can be measured with good precision for single objects. In this case, the observed $y$ are randomly distributed around $Y$ assuming a constant statistical uncertainty of $\delta y=0.1/\ln(10)$. Since data samples of WL cluster masses consist usually of a few dozens, we consider the fitting of a random subset of $10^2$ fictitious data points or a sample of measured $y$ of the same size. 

Results are summarised in Table~\ref{tab_summary_single} and Fig.~\ref{tab_PDF_stacking}, where we compare results when the values of $y$ are either directly observed for individual clusters (`$y$ observed') or recovered from the stacked data (`$y$ recovered'). The reconstruction method can recover the intrinsic parameters but with larger statistical uncertainties than the ideal fitting to observed data, even though this estimated precision is driven by our arbitrary choice for the statistical uncertainties $\delta y$ and  $\delta \bar{y}$. The smaller the statistical uncertainties, the better the precision which the scatter can be recovered to.

\section{A test case: the AMICO-KiDS clusters}
\label{sec_amico_kids}

\begin{table}
\caption{Parameters of the conditional scaling relation (WL mass given optical richness) for the AMICO-KiDS-DR3 clusters.}
\centering
	\begin{tabular}[c]{l r@{$\,\pm\,$}l}
	\hline
 \noalign{\smallskip}  
	 parameter 	& \multicolumn{2}{c}{observed}  \\ 
	 \hline
	\noalign{\smallskip}  
$\alpha_{m_\text{WL}|\lambda_*}$      	&	0.00 	&	0.04	\\
$\beta_{m_\text{WL}|\lambda_*}$        	&	1.69 	&	0.08	\\
$\gamma_{m_\text{WL}|\lambda_*}$      	&	-0.94	&	0.60	\\
$\log(\sigma_{m_\text{WL}|\lambda_*})$	&	-1.13	&	0.53	\\
\hline
\end{tabular}
\label{tab_amico_kids_YIX}
\end{table}

\begin{figure*}
\centering
\includegraphics[width=\hsize]{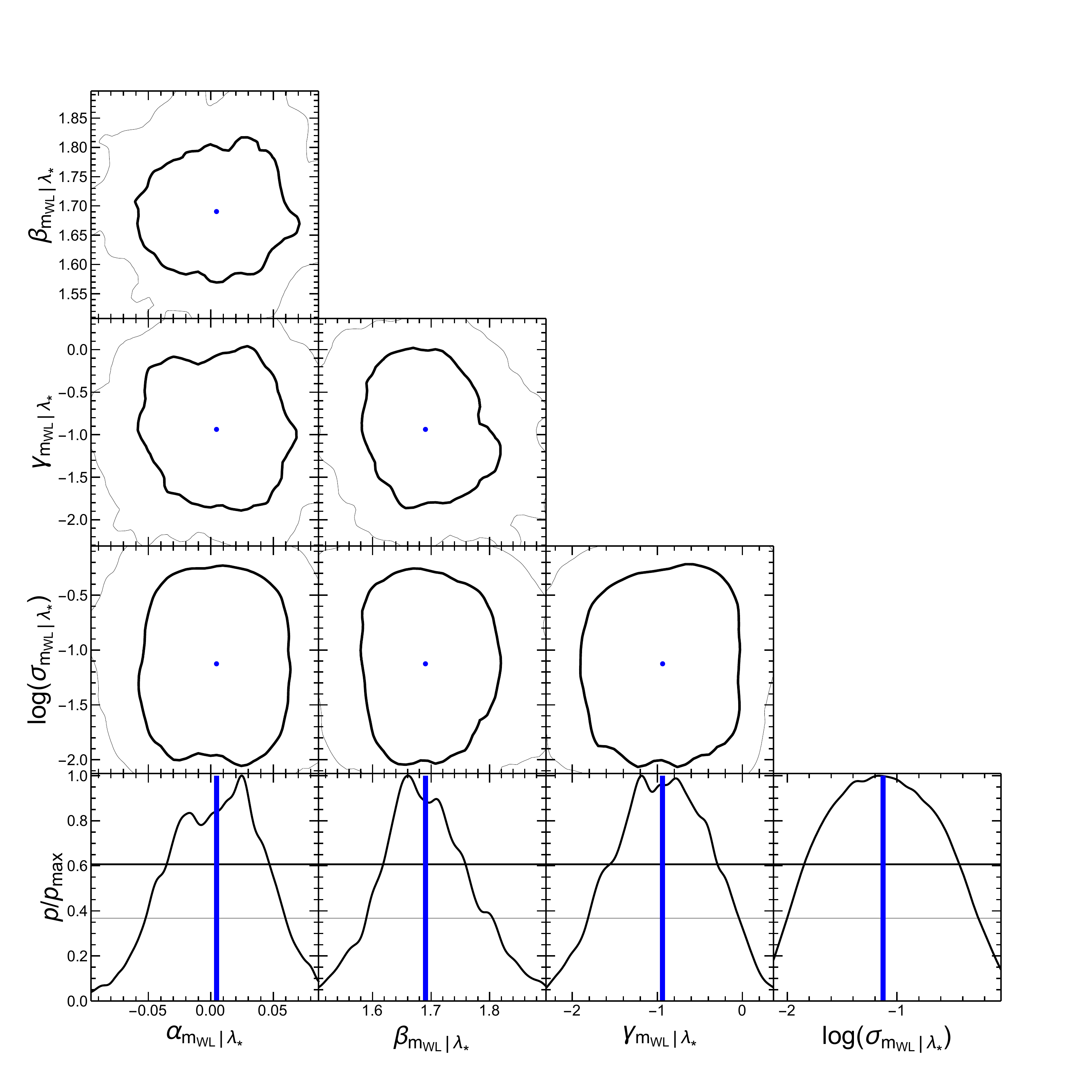} 
\caption{Probability distributions of the parameters of the scaling relation of the weak lensing mass for a given richness of the clusters in the KiDS-DR3 AMICO catalog. The intercept, slope, redshift-evolution, and intrinsic scatter are denoted as $\alpha$, $\beta$, and $\gamma$, respectively. The thick and thin black contours include the 1-$\sigma$ and 2-$\sigma$ confidence regions in two dimensions, here defined as the regions within which the probability is larger than $\exp (-2.3/2)$ and $\exp(-6.17/2)$ of the maximum, respectively. The bottom row shows the marginalised 1D distributions, renormalised to the maximum probability. The thick and thin black horizontal lines denote the confidence limits in one dimension, i.e. $\exp(-1/2)$ or $\exp(-4/2)$ and of the maximum. The blue symbols mark the biweight estimator.}
\label{fig_amico_kids_2PDF_YIX.pdf}
\end{figure*}

\begin{table}
\caption{Parameters of the scaling relations for the AMICO-KiDS-DR3 clusters when both the WL mass and the optical richness are considered as scattered proxies of a latent $Z$ variable.}
\centering
	\begin{tabular}[c]{l r@{$\,\pm\,$}l}
	\hline
 \noalign{\smallskip}  
	 parameter 	& \multicolumn{2}{c}{observed}  \\ 
	 \hline
	\noalign{\smallskip}  
$\alpha_{m_\text{WL}|Z\lambda_*}$      	&	0.00 	&	0.01	\\
$\beta_{m_\text{WL}|Z\lambda_*}$        	&	1.70 	&	0.05	\\
$\gamma_{m_\text{WL}|Z\lambda_*}$      	&	$-0.97$	&	0.22	\\
$\log(\sigma_{m_\text{WL}|Z\lambda_*})$	&	$-0.63$	&	0.01	\\
$\log(\sigma_{\lambda_*|Z\lambda_*})$  	&	$-1.63$	&	0.24	\\
\hline
\end{tabular}
\label{tab_amico_kids_YIZ}
\end{table}

\begin{figure*}
\centering
\includegraphics[width=\hsize]{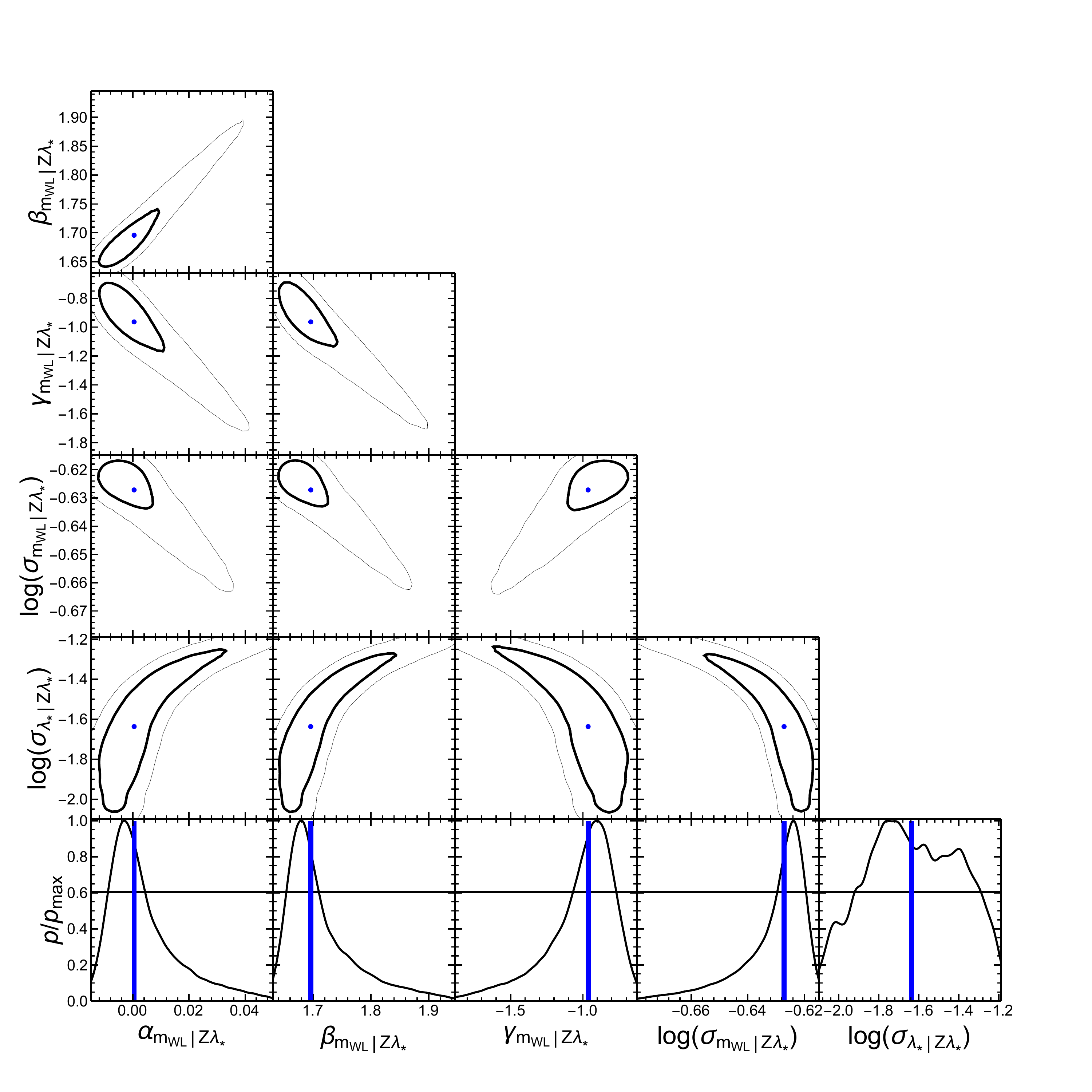} 
\caption{Probability distributions of the parameters of the scaling relation of the weak lensing mass and richness for a given unscattered richness of the clusters in the KiDS-DR3 AMICO catalog. The intercept, slope, redshift-evolution, and intrinsic scatters are denoted as $\alpha$, $\beta$, $\gamma$, and $\sigma$, respectively. The thick and thin black contours include the 1-$\sigma$ and 2-$\sigma$ confidence regions in two dimensions, here defined as the regions within which the probability is larger than $\exp (-2.3/2)$ and $\exp(-6.17/2)$ of the maximum, respectively. The bottom row shows the marginalised 1D distributions, renormalised to the maximum probability. The thick and thin black horizontal lines denote the confidence limits in one dimension, i.e. $\exp(-1/2)$ or $\exp(-4/2)$ of the maximum. The blue points and the blue vertical lines mark the biweight estimator.}
\label{fig_amico_kids_2PDF_YIZ.pdf}
\end{figure*}

We apply our procedure to the catalog of galaxy clusters detected with the optimal filtering algorithm AMICO \citep{bel+al18,mat+al19} in the sky area covered by the KiDS Data Release 3 \citep{dej+al13,kui+al15,dej+al17}. KiDS-DR3 covers $\sim440\deg^2$ in four optical bands $u$, $g$, $r$, $i$, down to the limiting magnitudes (calculated as $5\sigma$ in a 2\arcsec aperture) of 24.3, 25.1, 24.9, and 23.8, respectively \citep{dej+al17} . 

The catalogue comprises 8092 candidate clusters at redshifts $z < 0.8$ \citep{rad+al17,mat+al19}. The 6961 objects in the redshift range $0.1 < z < 0.6$ were mass calibrated in \citet{bel+al19}, who performed a WL stacked analysis by binning clusters according to redshift and two different mass proxies, namely the amplitude $A$ (the returned signal amplitude of the optimal filtering procedure) and the richness $\lambda_*$ (the sum of membership probabilities). The mass range of the detected clusters extends over more than one order of magnitude, down to $M_{200}\la 10^{13}M_\odot$. In the following, we exploit the WL mass estimates of \citet{bel+al19}, $M_\text{WL}$, but we extend the analysis of the mass--observable relation to estimate the intrinsic scatter.

Following the notation of the previous sections, we define the scattered proxies $X$ and $Y$ as
\begin{eqnarray}
X & = & \log(\lambda_*/30), \\
Y & = & \log(M_\text{WL}/M_\odot/10^{14}), \\
Z & = & \log(Z_{\lambda_*}/30)
\end{eqnarray}
where $Z$ is the unscattered latent variable, which $X$ and $Y$ are scattered proxies of. The property $Z_{\lambda_*}$ can be thought of as the richness we would measure if there was no intrinsic scatter in the true mass-richness relation. We consider $X$ as unbiased, i.e. we fix the scaling parameters to $\alpha_{X|Z}=0$ and $\beta_{X|Z}=1$. The variables $X$ and $Y$ differ from their measured values $x$ and $y$ for statistical uncertainties or systematic errors, see App.~\ref{sec_syst}. We first fit the stacked relation in order to infer the estimated WL mass of a cluster given its richness. Clusters are grouped in bins of approximately equal SNR, with the number of clusters per bin ranging from more than one thousand for the low richness bins, to a few dozens for the large richness bins \citep[table 3]{bel+al19}. In addition to the statistical uncertainty on the estimation of the stacked WL mass, we consider a systematic error of $7.6$ per cent due to impure selection of background galaxies, photometric redshifts estimates, shear measurements, projection effects, and halo modelling \citep{bel+al19}.

Projection effects or orientation bias can play a major role both in WL mass estimates and cluster detection. The processes of optical cluster selection and richness estimation can be biased, leading to stacking of structures that are preferentially elongated along the line of sight. WL masses can be then overestimated \citep{die+al14}. These effects can make the intrinsic scatters of WL mass and richness at a given true mass correlated. Unfortunately, degeneracy effects, see Sec.~\ref{sec_scat}, prevent a full recovery of the scatter correlation, whose analysis would need the 
joint comparison of multiple proxies \citep{far+al19,ser+al20_hscxxl}. In the following, we neglect the correlation between the intrinsic scatters of WL mass and richness.

The expected value of $Y$ given $X$ is expressed as \citepalias{se+et15_comalit_IV},
\beq
\label{eq_bug_multi_1}
\langle Y|X \rangle  =  \alpha_{Y|X}+\beta_{Y|X} X + \gamma_{Y|X} \log F_z ,
\eeq
where $\alpha$ denotes the normalisation, the slope $\beta$ accounts for the dependence on $Z$, and the slope $\gamma$ accounts for the redshift evolution. $F_z$ is the renormalised Hubble parameter, $F_z=E_z/E_z(z_\text{ref})$. For the AMICO-KiDS-DR3 sample, we fix $z_\text{ref}=0.35$, close to the mean redshift of the full sample. At a given $X$, $Y$ is distributed around $X$ with a scatter $\sigma_{Y|X}$, which we assume to be constant. 

Results are summarised in Tab.~\ref{tab_amico_kids_YIX} and Fig.~\ref{fig_amico_kids_2PDF_YIX.pdf}. The scaling parameters are in agreement with \citet{bel+al19}. The large number of clusters per bin makes the effective scatter small, so that concurring scaling parameters $\alpha$ and $\beta$ can be recovered notwithstanding the fitting method. In particular, \citet{bel+al19} did not have to consider the scatter as a model parameter. However, thanks to the regression procedure described here, we can fit the intrinsic scatter too. We find that the richness is an excellent mass proxy. The intrinsic scatter of the WL mass of a single cluster at a given richness is $\sigma_{m_\mathrm{WL}|\lambda_*}=18\pm22$ per cent, with a marginalised probability distribution in linear space $p(\sigma_{m_\mathrm{WL}|\lambda_*})$ peaked at very low values and with an extended tail at large values. The probability that the scatter is lower than 10 (or 5) per cent is $\sim 35$ (or 17) per cent. Since the posterior probability distribution is skewed with an extend tail, the logarithm (in base 10) of the biweight estimator of the conditional scatter ($\log( \text{CBI} [ \sigma_{m_\mathrm{WL}|\lambda_*} ] )\sim-0.7$) is significantly larger than the biweight estimator of the logarithms ($\text{CBI} [ \log(\sigma_{m_\mathrm{WL}|\lambda_*}) ]\sim-1.1$), see Table~\ref{tab_amico_kids_YIX}).

We then reconstruct the scattered distribution of the WL masses based on the richness distribution. Results are summarised in Tab.~\ref{tab_amico_kids_YIZ} and Fig.~\ref{fig_amico_kids_2PDF_YIZ.pdf}. We find a low value for the intrinsic scatter of the richness, $\log(\sigma_{\lambda_*|Z\lambda_*})\sim-1.6$, which makes the slope $\beta_{Y|Z}$ similar to $\beta_{Y|X}$. Statistical uncertainties on the measured richness reported in the catalog are of the order of $\sim 20$ per cent. They are estimated with the analysis of mock galaxy catalogues derived directly from the data to fully reproduce their statistical properties including photo-$z$ uncertainties, unknown absorption across the survey, missing data, spatial correlation of galaxies and galaxy clusters \citep{mat+al19}. As a result, the formal statistical uncertainty accounts for projection effects too, which are one of the main source of dispersion. If this major contribution is treated as a source of statistical uncertainty, it does not contribute to the intrinsic scatter of the richness, which we find to be small.

\section{Mass proxies}
\label{sec_prox}

Based on the AMICO richness, the mass of the clusters in the KiDS-DR3 can be determined to a $\sim20$ per cent precision. This result cannot be compared to performances of other richness based  proxies. The richness somehow counts the number of galaxies in a cluster but its definition depends on the measurement process. We can count galaxies in different magnitude ranges and aperture radii; we can look for red-sequence galaxies or galaxies with similar photometric redshifts. Furthermore, the performance of a proxy optimised on a calibration sample of well selected clusters with high quality data can be better than for the very numerous candidate clusters found in a very large and shallow survey.

Even if a fair comparison cannot be performed, it can be still useful to review the performances of some richness estimators as mass proxy. \citet{wen+al12} identified overdensities of galaxies around the brightest cluster galaxies (BCGs) through their photometric redshifts. The optical richness is defined as the ratio of the total $r$-band luminosity within an empirically determined radius and the evolved characteristic galaxy luminosity. Based on a collection of 1191 clusters with masses estimated with either X-ray or SZ proxies, they found that the mass of the 132684 candidate galaxy clusters detected in the SDSS (Sloan Digital Sky Survey) DR12 can be estimated with a scatter of $\sim 40$ per cent \citep{we+ha15}.

\citet{ryk+al12} considered a red-sequence-matched filter richness estimator implemented on the maxBCG cluster catalog. Using the X-ray luminosity from the ROSAT All-Sky Catalog as mass proxy, they found a scatter in mass at fixed richness of $\sim 20-30$ per cent depending on the richness, and comparable to that for total X-ray luminosity. 

The red-sequence Matched-filter Probabilistic Percolation (redMaPPer) algorithm is a photometric cluster finding algorithm which identifies galaxy clusters as overdensities of red-sequence galaxies \citep{ryk+al14}. \citet{ro+ry14} evaluated the performance of the Sloan Digital Sky Survey (SDSS) DR8 redMaPPer photometric cluster catalog by comparison to overlapping X-ray and SZ-selected catalogs from the literature. Based on the X-ray temperature-richness and gas mass-richness relations, they estimated a mass scatter of $\sim25$ per cent.

The CAMIRA (Cluster finding Algorithm based on Multi-band Identification of Red-sequence gAlaxies) algorithm is a red-sequence cluster finder based on a stellar population synthesis models \citep{ogu14}. \citet{mur+al19} adopted a forward modelling approach to fit the abundance and stacked lensing profiles of the CAMIRA clusters detected in the Hyper Suprime-Cam (HSC) survey first-year data. 
They found that the scatter values of the mass at a given richness for the Planck model ($\sim 30$ per cent) are systematically larger than those for the WMAP model. They also found that the scatter values for the Planck model increase toward lower richness values, whereas those for the WMAP model are consistent with constant values as a function of richness.

\section{Conclusions}
\label{sec_conc}

The potential of galaxy cluster number counts as cosmological probe can be fully exploited if the statistical properties of the sample are well characterised and if the mass calibration is accurate. In present and planned surveys, investigators have shown confidence that the completeness and purity of selected clusters can be well measured \citep{euclid_ada+19}. Uncertain mass calibration has been the designated scapegoat for inconclusive results \citep{planck_2015_XXIV,des_abb+al20}.  A proper treatment of scaling relation and mass calibration is then crucial to settle the question. WL masses are regarded as the most reliable mass estimates. Stacking enable us to calibrate the observable--mass relation down to the very low mass haloes discovered by large and deep surveys. This technique helps in studying the scaling parameters without extrapolation but can make some parameter estimations problematic.
The intrinsic scatter should be derived from the data as well but it is usually constrained through strong priors, which could bias the cosmological inference if misplaced. In this paper, we have proposed a Bayesian method to infer the intrinsic scatter from stacked observable--mass relations. 

Bayesian inference is a solid tool to infer unbiased physical quantities in problems with a large number of manifest or latent variables and parameter degeneracy. In the simplest case of uncorrelated data, the intrinsic scatter of the stacked signal from $N_\text{stack}$ clusters scales as $N_\text{stack}^{-1/2}$ of the scatter of individual objects. However, some sources of scatters can be correlated and the variance of a cluster stack does not scale simply as $1/N_\text{stack}$. For example, the positions of galaxy clusters are correlated and the variance in the stacked WL signal due to uncorrelated structure decreases somewhat less steeply than $1/N_\text{stack}$ \citep{mcc+al19}. In this case the scaling of the stacked scatter has to be properly weighted.

Whereas targeted observations are very expensive and feasible only for relatively small data samples \citep{wtg_I_14,pos+al12,ste+al20}, mass proxies based on optical richness are cheap by design in large surveys and can provide accurate and precise masses even for small groups. As a test case, we applied our approach to the AMICO clusters in the KiDS survey. The method showed that the optical richness determined by the AMICO algorithm itself is a reliable mass proxy, with a scatter of $\sim 20$ per cent. This is comparable to the precision attainable with direct WL or X-ray mass measurements for very deep observations \citepalias{se+et15_comalit_I}. 

The knowledge of the observable--cluster mass scaling relation is crucial to fulfil the potential of galaxy clusters as cosmological probes. Thanks to strong constraints on scatter and mass bias, constraints on dark energy from analyses of number counts and clustering can be significantly improved. \citet{sar+al16} showed that for an Euclid-like survey the figure of merit for the parameters of the dark energy equation of state increases by a factor of $\sim 4$ if the parameters of the scaling relation are accurately known. Precision cosmology requires that the scaling parameters and the scatter of the scaling relation are determined together with the cosmological parameters \citep{mur+al19}.

\section*{Acknowledgements}
SE and MS acknowledge financial contribution from contract ASI-INAF n.2017-14-H.0 and INAF `Call per interventi aggiuntivi a sostegno della ricerca di main stream di INAF'. FM and LM acknowledges support from grants ASI n.I/023/12/0, ASI-INAF n.2018-23-HH.0, PRIN MIUR 2015 `Cosmology and Fundamental Physics: illuminating the Dark Universe with Euclid', and PRIN-MIUR 2017 WSCC32.

This research has made use of NASA's Astrophysics Data System (ADS) and of the NASA/IPAC Extragalactic Database (NED), which is operated by the Jet Propulsion Laboratory, California Institute of Technology, under contract with the National Aeronautics and Space Administration.

\section*{Data availability}
The software \texttt{LIRA} (LInear Regression in Astronomy) is publicly available from the Comprehensive R Archive Network at \url{https://cran.r-project.org/web/packages/lira/index.html}.

The data underlying this article will be shared on reasonable request to the corresponding author.


\appendix

\section{Bivariate normal distribution}
\label{sec_biva}

Let $X$ and $Y$ be two scattered proxies of $Z$. The marginalised bivariate normal distribution of $X$ and $Y$ can be written as,
\beq
p(X,Y) = {\cal N}^\mathrm{(2)} (\left\{ X, Y \right\} | \left\{ \mu_X, \mu_Y \right\}, \Sigma_{XY} ),
\eeq
where ${\cal N}^\mathrm{(2)}$ is the bivariate Gaussian distribution, the mean values of $X$ and $Y$ are
\begin{eqnarray}
\mu_X & = & \alpha_{X|Z} + \beta_{X|Z} \mu_Z ,  \\
\mu_Y & = & \alpha_{Y|Z} + \beta_{Y|Z} \mu_Z ,
\end{eqnarray}
respectively, and the covariance matrix $\bmath{\Sigma}_{XY}$ can be expressed as
\begin{equation} 
\label{eq_cov}
  \bmath{\Sigma}_{XY} = \left(\begin{array}{cc}
      \sigma^2_{X} & \rho_{XY}\sigma_X \sigma_Y  \\
       \rho_{XY}\sigma_X \sigma_Y & \sigma^2_{Y} \\
    \end{array}\right),
\end{equation}
with
\begin{eqnarray}
\sigma_X^2  & = & \sigma_{X|Z}^2 + \beta_{X|Z}^2 \sigma_Z^2\, ,\\
\sigma_Y^2  & = & \sigma_{Y|Z}^2 + \beta_{Y|Z}^2 \sigma_Z^2\, , \\
\rho_{XY}     & = &\frac{1}{
\left(1+ \frac{\sigma_{X|Z}^2}{\beta_{X|Z}^2 \sigma_Z^2} \right)^{1/2} 
\left(1+ \frac{\sigma_{Y|Z}^2}{\beta_{Y|Z}^2 \sigma_Z^2}\right)^{1/2}  
}\, .
\end{eqnarray}

The probability of $X$ and $Y$ can be also written in terms of the conditional probability of $Y$ given $X$ thanks to the chain rule,
\beq
\label{eq_Y|X_1}
p(X,Y) ={\cal N}(Y | \alpha_{Y|X}+  \beta_{Y|X} X, \sigma_{Y|X}) {\cal N}(X |\mu_X,\sigma_{X}) \, ,
\eeq
where
\begin{eqnarray}
\alpha_{Y|X} & = & \mu_Y - \frac{\beta_{Y|Z}}{\beta_{X|Z} } \frac{ \mu_X}{1 + \frac{\sigma_{X|Z}^2}{\beta_{X|Z}^2 \sigma_Z^2} }\, ,  \label{eq_Y|X_2} \\
\beta_{Y|X}   & = & \frac{\beta_{Y|Z}}{\beta_{X|Z} }\frac{1}{1 + \frac{\sigma_{X|Z}^2}{\beta_{X|Z}^2 \sigma_Z^2} }\, , \label{eq_Y|X_3} \\
\sigma_{Y|X}^2 & = & \sigma_{Y|Z}^2 + \frac{\beta_{Y|Z}^2}{\beta_{X|Z}^2}\frac{\sigma_{X|Z}^2}{1 + \frac{\sigma_{X|Z}^2}{\beta_{X|Z}^2 \sigma_Z^2} }\, . \label{eq_Y|X_4}
 \end{eqnarray}

The normalisation and the scatter can be rewritten in a more compact form in terms of the slope $\beta_{Y|X}$ as
\begin{eqnarray}
\alpha_{Y|X} & = & \mu_Y - \beta_{Y|X} \mu_X , \\
\sigma_{Y|X}^2 & = & \sigma_{Y|Z}^2 + \beta_{Y|X} \frac{\beta_{Y|Z}}{\beta_{X|Z}}\sigma_{X|Z}^2 .
 \end{eqnarray}
 
The probability of $X$ given $Y$ can be obtained from the above expression by inverting $X$ and $Y$.

\section{Systematic errors}
\label{sec_syst}



The measured $x$ and $y$ and the latent values $X$ and $Y$ are related as
\begin{multline}
P(x_i,y_i | X_i, Y_i)  \propto  \label{eq_bug_multi_2}\\
 \ {\cal N}^\text{2}\left(\{X_i-\delta x_\text{syst},Y_i-\delta y_\text{syst} \},\mathbf{V}_{\delta,i}\right)  \times {\cal H}(y_{\text{th},in}),
\end{multline}
where ${\cal H}$ is the Heaviside function, $\mathbf{V}_{\delta,n}$ is the covariance matrix of the $i$-th cluster accounting for statistical uncertainties, and $\delta x_\text{syst}$ and $\delta y_\text{syst}$ are systematic uncertainties which affect all clusters in the same way.


The probability distribution is truncated for $y_{in}<y_{\text{th},in}$ to correct for the Malmquist bias if only clusters above the observational thresholds (in the response variables) are included in the sample \citepalias{ser+al15_comalit_II}.

\section{Reproducibility of the results}
\label{sec_repr}


To allow the reproducibility of our results, we provide the commands used in Sec.~\ref{sec_amico_kids}.  Let \texttt{x} and \texttt{y}, \texttt{delta.x} and \texttt{delta.y}, \texttt{covariance.xy}, and \texttt{z} be the vectors storing the values of the observed $\bmath{x}$ and $\bmath{y}$, their uncertainties $\bmath{\delta_x}$ and $\bmath{\delta_y}$, the uncertainty covariances $\bmath{\delta_{xy}}$, and the redshifts $\bmath{z}$, respectively. If not stated otherwise, priors and parameter values are set to default.

\begin{itemize}
\item For regressions of stacked data, without scatter on the $X$ variable, the analysis is performed with the command

\noindent \texttt{> mcmc <- lira (x, y, delta.x = delta.x, delta.y = delta.y, delta.y.syst='dnorm(0.0,(0.076/log(10.))\^{}-2)', z = z, z.ref = 0.35, gamma.mu.Z.Fz=0.0, gamma.sigma.Z.D='dt$'$, n.chains = 4, n.adapt = 5*10\^{}3, n.iter = 5*10\^{}4) },

\noindent where the covariate distribution is modelled as a Gaussian function with redshift evolving mean and standard deviation (\texttt{gamma.sigma.Z.D=$'$dt$'$}). Each of the \texttt{n.chains = 4} chain was \texttt{n.iter = 5$\times$10$^4$} long, and the number of iterations for inizialisation was set to \texttt{n.adapt =5*10$^3$}. The prior on the systematic error on $y$ is modelled as a zero centred Gaussian with standard deviation of $0.076/\log(10.)$.
 
\item For regressions with scatter on both the $Y$ and the $X$ variables, the analysis is performed with the command

\noindent \texttt{> mcmc <- lira (x, y, delta.x = delta.x, delta.y = delta.y, covariance.xy = covariance.xy, z = z, z.ref = 0.35, sigma.XIZ.0 = $'$prec.dgamma$'$, gamma.mu.Z.Fz=0.0, gamma.sigma.Z.D='dt$'$, n.chains = 4, n.adapt = 5*10\^{}3, n.iter = 5*10\^{}4) },

\noindent where the argument \texttt{sigma.XIZ.0 = $'$prec.dgamma$'$} makes the scatter in $X$ a parameter to be fitted with a prior on the precision described by a Gamma distribution.
\end{itemize}

\end{document}